\documentclass{psp-book9x6-modified}

\usepackage{psp-book-van-modified} 
\usepackage{cite,graphicx,color,amsmath,booktabs,microtype,afterpage,subfigure,fixmath,truncate}
\usepackage{mathpazo} 
\usepackage[scaled]{helvet} 
\usepackage{bibunits} 
\usepackage{perpage} 
\usepackage[colorlinks,plainpages=false,linkcolor=black,urlcolor=black,citecolor=black,pdfpagemode=UseNone,pdfstartview=FitBH]{hyperref}
\usepackage[dvipsnames]{xcolor}
\usepackage[T1]{fontenc}\usepackage[latin1]{inputenc}

\let\Gamma\varGamma
\let\Delta\varDelta
\let\Theta\varTheta
\let\Lambda\varLambda
\let\Xi\varXi
\let\Pi\varPi
\let\Sigma\varSigma
\let\Upsilon\varUpsilon
\let\Phi\varPhi
\let\Psi\varPsi

\makeatletter
\renewcommand\thefigure{\thechapter.\@arabic\c@figure}
\makeatother

\usepackage{perpage}\MakePerPage{footnote}

\defaultbibliographystyle{inosov-book}

\copyline{Rare-Earth Borides}{Dmytro S. Inosov (ed.)}{2020}
\title{Rare-Earth Borides} 

\makeindex

\graphicspath{{Figures/}} 

\begin{document}









\cleardoublepage
\setcounter{page}{1}
\setcounter{chapter}{0}












\chapter{\bf \mbox{Bulk and surface properties of SmB$_\text{6}$}\label{Chapter:FiskRosa}}
\addtocontents{toc}{\vspace{-2pt}\hspace{2.7em}\textit{by Priscila~F.~S.~Rosa and Zachary~Fisk}\smallskip}

\chapauth{Priscila F.~S.~Rosa$^{\text{a},\ast}$ and Zachary Fisk$^\text{b}$
\chapaff{\noindent
$^\text{a}$Los Alamos National Laboratory, Los Alamos NM 87545, U.S.A.\\
$^\text{b}$University of California at Irvine, Irvine CA, U.S.A.\\
$^\ast$E-mail address: \href{mailto:pfsrosa@lanl.gov}{pfsrosa@lanl.gov}}}

\begin{bibunit}

\section*{Abstract}
\label{Fisk-abstract}

Samarium hexaboride crystallizes in a simple cubic structure (space group \#221, $Pm\overline{3}m$), but its properties are far from being straightforward. Initially classified as a Kondo insulator born out of its intriguing intermediate valence ground state, SmB$_6$ has been recently predicted to be a strongly correlated topological insulator. The subsequent experimental discovery of surface states has revived the interest in SmB$_6$, and our purpose here is to review the extensive and in many aspects perplexing experimental record of this material. We will discuss both surface and bulk properties of SmB$_6$ with an emphasis on the role of crystal growth and sample preparation. We will also highlight the remaining mysteries and open questions in the field.\clearpage

\section{Introduction}
\label{Fisk-intro}

First synthesized in polycrystalline form in 1932, SmB$_6$ started to be more heavily studied only in the 1970s owing to the successful growth of single crystals, the prospect of its use in thermoionic emission applications, and its intriguing mixed valence (Sm$^{2.6+}$), which gives rise to a Kondo-insulating ground state. The properties of SmB$_6$ have been investigated by a wide variety of experimental techniques and have been reported in more than 800 scientific articles, with about half of them being published just in the past decade. This flourish of research was motivated, to a great extent, by theoretical predictions invoking topological concepts recently extended to condensed-matter physics. SmB$_6$ remains a puzzling material, and here we take the challenge of reviewing our current knowledge of its bulk and surface properties.

This book chapter is organized as follows. First, we will discuss structural aspects and crystal growth methods typically used to synthesize single crystals of SmB$_6$. We will then present a very brief summary of proposed theoretical models. The third and fourth sections will review surface and bulk properties of SmB$_6$, respectively.

\section{Crystal growth and structural properties}\index{SmB$_6$!crystal growth|(}
\label{Fisk-growth}

An important message of this chapter is that the physical properties of SmB$_6$ are sensitive to growth conditions. It will become clear in the next sections that advances in materials characterization have made the investigation of single crystals the current accepted norm, but we note that SmB$_6$ was first synthesized in polycrystalline form via borothermal reduction \cite{Fis_Synthesis-1932}. Typically, a homogeneous mixture of samarium oxide (Sm$_2$O$_3$) and boron powder is pressed into a pellet and heated by induction to temperatures from 1500 to 1800\,$^\circ$C in vacuum or in a hydrogen atmosphere \cite{Fis_osti_4141572}.

For the growth of single-crystalline SmB$_6$, molten flux\index{SmB$_6$!crystal growth!Al flux method} and floating-zone methods are the typical methods of choice. The molten-flux technique often allows the use of a suitable low-melting-point element that acts as a solvent out of which crystals can be grown at much lower temperatures than the melting point of the compound \cite{Fis_Fisk1989, Fis_Canfield1992, Fis_Rosa2019}. This is particularly true here for SmB$_6$, and aluminum is the flux of choice. Though the solubility of hexaborides in molten Al is quite low in general, melts with typical concentration of the hexaboride of order $10^{-3}$ molar make the growth of millimeter-sized single crystals possible. The synthesis is performed in alumina tube furnaces at temperatures from 1500 to 1150\,$^\circ$C in a protective atmosphere of ultra-high-purity argon, which flows at a slow rate. Usually, 50~ml alumina crucibles hold the melts, and chemical etching with NaOH solution is used to remove the Al at room temperature. A more benign leaching can be accomplished by growing from an Al-Ga melt, which can be leached using H$_2$O. Independent of the etching procedure, the etchant does not attack the hexaboride; however, larger crystals often enclose Al lamellae which cannot be removed by etching. This is the main disadvantage of synthesizing SmB$_6$ out of flux. As we will discuss below, x-ray computed tomography experiments on flux-grown crystals reveal the presence of co-crystallized, epitaxial aluminum. Nevertheless, aluminum inclusions can be mechanically removed by polishing.
\index{SmB$_6$!crystal growth!floating-zone method}

The traditional optical-image floating zone growth technique is based on the directional solidification of a crystal from a liquid floating zone heated by bulb lamps and parabolic mirrors \cite{Fis_Prokofiev2019}. The elimination of an alumina crucible and of the aluminum flux prevents some of the issues with the flux technique; however, crystals grown by the floating zone method require much higher temperatures ($T>2000$\,$^\circ$C), which lead to composition variations. Powder x-ray diffraction in consecutive cuts of a floating-zone grown crystal reveals a systematic change in lattice parameters \cite{Fis_Phelan2016}. Precisely identifying the origin of the composition change (e.g. Sm/B vacancies, crystallographic defects, etc.) is challenging and remains a major open question in the field.

\begin{figure*}[!t]
\includegraphics[width=\textwidth]{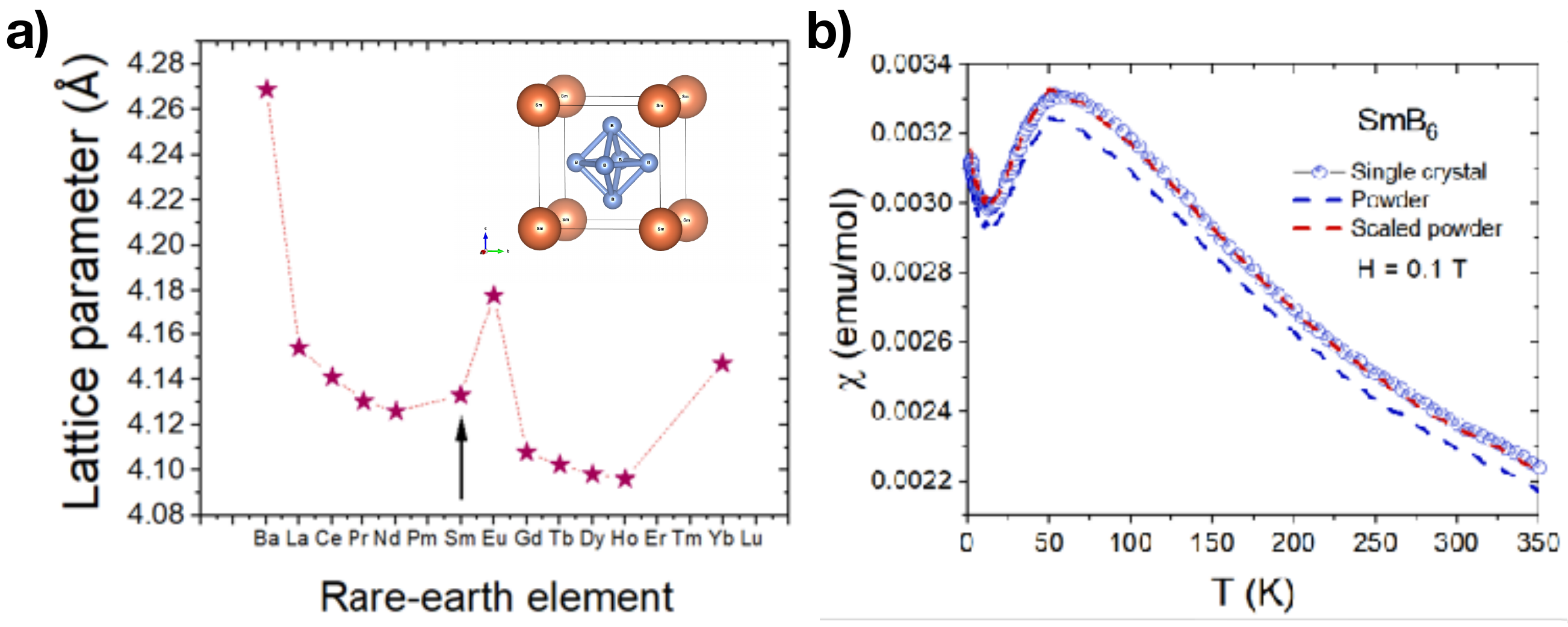}
\caption{(a)~Lattice parameters of the cubic hexaborides $R$B$_6$ ($R$ is a rare-earth element) at room temperature. Inset shows the cubic unit cell. (b)~Temperature dependent magnetic susceptibility measured on an Al-flux grown single crystal as well as its powder. The scaling factor is $1.05$ to account for a small loss of powder mass.}
\label{fig:FigA-LatticeParameters}
\end{figure*}

The simple CsCl-type arrangement of metal atoms and B$_6$ octahedra found in the rare-earth and alkaline-earth hexaborides was known from the early x-ray studies \cite{Fis_Allard1932, Fis_Pauling1934}. The electronic closed-shell configuration for a B$_6$ octahedron in this structure was determined later by molecular cluster calculations~\cite{Fis_Longuet1954}. Twenty electrons were found to be required for the closed-shell configuration of an octahedron. This finding suggested that divalent metals forming hexaborides should be semiconducting, whereas trivalent ones should be metallic. Subsequent de~Haas\,--\,van Alphen\index{SmB$_6$!quantum oscillations} studies of metallic LaB$_6$ were consistent with this picture, finding a Fermi surface containing one electron per formula unit \cite{Fis_Arko1976, Fis_Ishizawa1977}. Measurements of the ordinary Hall coefficient in PrB$_6$ and NdB$_6$ also found one electron per formula unit \cite{Fis_Onuki1989}.

SmB$_6$, however, is neither divalent nor trivalent. The intermediate valence\index{SmB$_6$!mixed valence} character of SmB$_6$ is already seen in the hexaboride lattice parameter variation across the rare-earth sequence shown in Fig.~\ref{fig:FigA-LatticeParameters}\,(a). The majority of rare-earth hexaborides is trivalent, whereas EuB$_6$ and YbB$_6$ are strictly divalent. Hexaborides with Er, Tm, and Lu do not form, possibly because their metallic radii are too small for the structure to be stable. Further evidence for the mixed valence comes from magnetic susceptibility measurements, shown in Fig.~\ref{fig:FigA-LatticeParameters}\,(b). At high temperatures, the magnetic susceptibility of SmB$_6$ falls between the values expected for Sm$^{2+}$ (Van Vleck) and for Sm$^{3+}$ ($J=5/2$ moment). Below about 60~K, the magnetic susceptibility decreases with decreasing temperature, in agreement with the opening of a gap. At temperatures below about 14~K, a Curie-like tail is observed. This tail is sample dependent, and measurements on a powdered single crystal indicate that it is a bulk effect [Fig.~\ref{fig:FigA-LatticeParameters}\,(b)].
\index{SmB$_6$!crystal growth|)}

\section{Theoretical remarks}
\label{Fisk-theory}

Studies of other intermediate valence rare-earth compounds with features similar to those seen in SmB$_6$ led to the suggestion of their classification as Kondo insulators, a group of semiconducting compounds containing $f$-elements which do not order magnetically \cite{Fis_Aeppli1992}. As a result, an isostructural semiconducting compound forming with non-$f$ elements has the integral valence of one of the $f$-configurations of the Kondo insulator. SmB$_6$ is seen as a prototypical Kondo insulator, the non-$f$ semiconducting counterparts being the alkaline-earth hexaborides. Though Kondo physics usually seems quite distinct from that of intermediate valence, in the case of SmB$_6$ it was found that dilute Sm impurities\index{impurities} in divalent SrB$_6$ showed a Kondo scale of 3~K, whereas dilute Sm impurities in trivalent LaB$_6$ appeared to be divalent, $4f^{6}$, with no Kondo behavior. This appears consistent with lattice parameter alloys studies of SmB$_6$/LaB$_6$ and SmB$_6$/YbB$_6$ in which Sm appears to go in as divalent and trivalent ions, respectively \cite{Fis_Kasaya1980}. Perhaps the most accepted picture that emerged for this class of small gap semiconducting materials is that the narrow gap arises from hybridization\index{hybridization gap} between the $4f$ electrons and $d$ conduction bands at the Fermi level \cite{Fis_Mott1974, Fis_Martin1979}. Other early scenarios include Wigner crystallization\index{Wigner crystallization} in the Kondo lattice \cite{Fis_Kasuya1979} and exciton-polaron models of charge fluctuations \cite{Fis_Kikoin1990}. In the former, carrier localization occurs owing to correlations. In the latter, an electronic instability leads to a mixed-valence state hosting a soft electronic exciton with mixed $4f$ and $5d$ wave functions. A more recent scenario argues for a mixed-valence\index{SmB$_6$!mixed valence} state originated from an unrecognized dynamical bonding effect, i.e., the coexistence of two Sm-B bonding modes corresponding to two different oxidation states driven by the motion of boron \cite{Fis_Robinson2019, Fis_Streltsov1999}.

In 2010, SmB$_6$ was proposed to be the first topological Kondo insulator,\index{topological Kondo insulator} i.e., a material that hosts a topologically protected surface state surrounding an insulating bulk driven by hybridization between Sm $f$ electrons and conduction $d$ electrons \cite{Fis_Dzero2010}. Band structure calculations indicate that the parity of the hybridized bands is inverted at three symmetry-equivalent $X$-points in the cubic Brillouin zone. The resulting topological invariant therefore predicts a non-trivial topological insulating state~\cite{Fis_Takimoto2011, Fis_Alexandrov2013, Fis_Lu2013}. Further, motivated by the unusual bulk properties we discuss below, more recent theories involve ingredients such as neutral Fermi surfaces,\index{Fermi surface!charge neutral} spin excitons,\index{spin exciton} fractionalized quasiparticles and disorder \cite{Fis_Batko2014, Fis_Baskaran2015, Fis_Zhang2016, Fis_Erten2016, Fis_Knolle2017, Fis_Erten2017, Fis_Ram2017, Fis_Chowdhury2018, Fis_Shen2018, Fis_Fuhrman2018, Fis_Pal2019, Fis_Skinner2019}. Our goal here is not to present a detailed overview of the numerous theories, but instead to present the ensemble of experimental data against which these theories need to be checked. For theoretical reviews, we refer the reader to \cite{Fis_Riseborough2000a, Fis_Dzero2016, Fis_Wu2017}.\vspace{-2pt}

\section{Surface properties}\label{Fisk-surface-sec}\vspace{-2pt}

\subsection{dc electrical conductivity}\index{SmB$_6$!dc electrical conductivity|(}
\label{Fisk-surface-dc}

\begin{figure*}[!b]
\includegraphics[width=\textwidth]{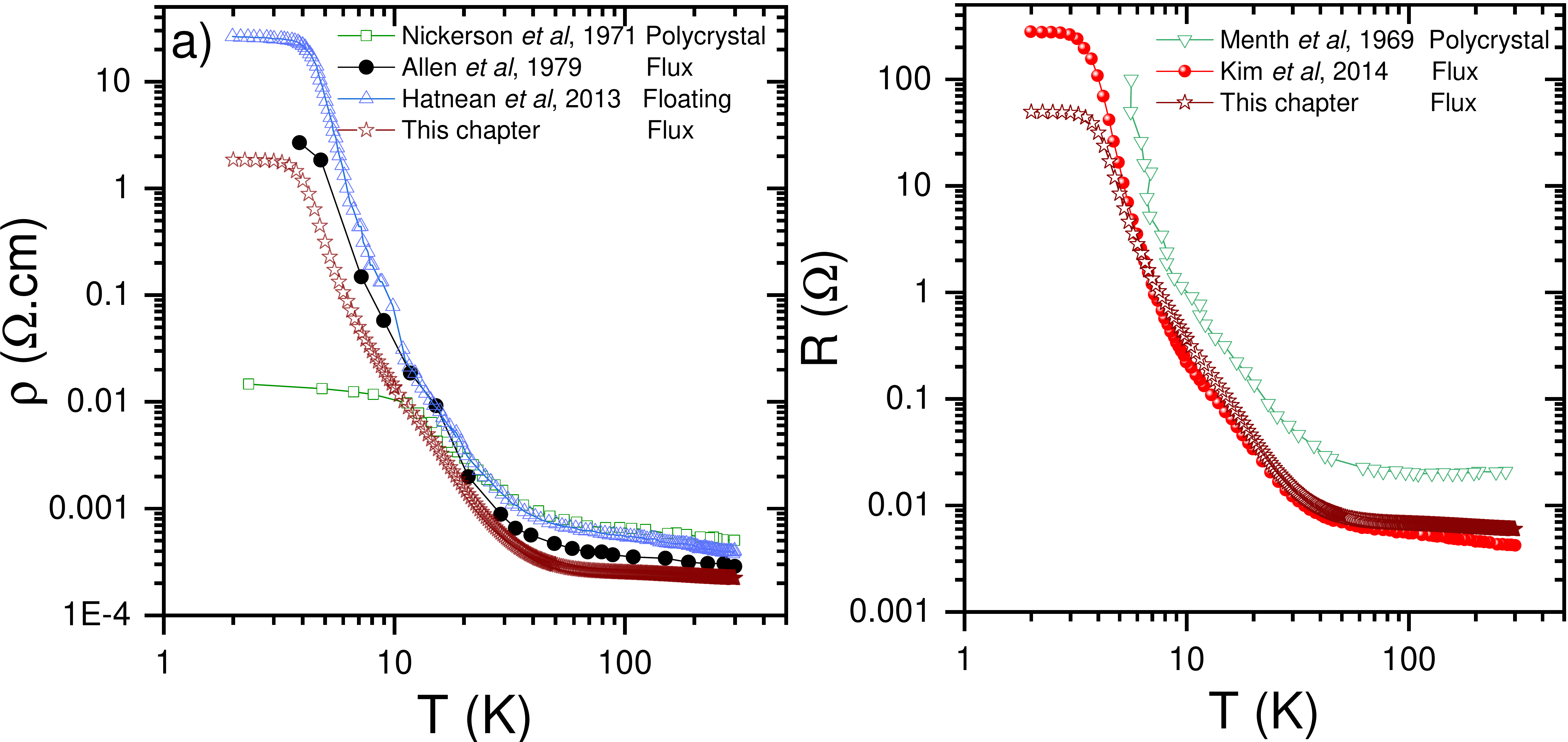}
\caption{Compilation of temperature dependent (a)~electrical resistivity and (b)~electrical resistance obtained in SmB$_6$ samples grown by different methods. Reproduced from Menth~\textit{et al.}~\cite{Fis_Menth1969}, Nickerson~\textit{et al.}~\cite{Fis_Nickerson1971}, Allen~\textit{et al.} \cite{Fis_Allen1979}, Hatnean~\textit{et al.} \cite{Fis_Hatnean2013} and Kim~\textit{et al.}~\cite{Fis_Kim2014}.}
\label{fig:Fig1-dc-resistance}
\index{SmB$_6$!electrical resistivity}
\end{figure*}

Figure~\ref{fig:Fig1-dc-resistance} displays a summary of representative dc electrical resistivity (panel a) and resistance (panel b) data of SmB$_6$ \cite{Fis_Menth1969, Fis_Nickerson1971, Fis_Allen1979, Fis_Hatnean2013, Fis_Kim2014}. These measurements are typically performed in a four-point configuration using a low-frequency ac excitation and a lock-in amplifier for detection.

At high temperatures, the electrical resistance of SmB$_6$ increases with decreasing temperature, in agreement with an insulating response. At around 15~K, a broad feature usually emerges, which we will come to later. Finally, at temperatures below about 4~K, the resistance saturates. This saturation was initially thought to be extrinsic and attributed to impurity conduction (e.g. holes from Sm vacancies); however, this plateau has been recently revisited due to the prediction of surface states.\index{surface states!in SmB$_6$}\index{SmB$_6$!surface states}

\begin{figure*}[!ht]
\includegraphics[width=\columnwidth]{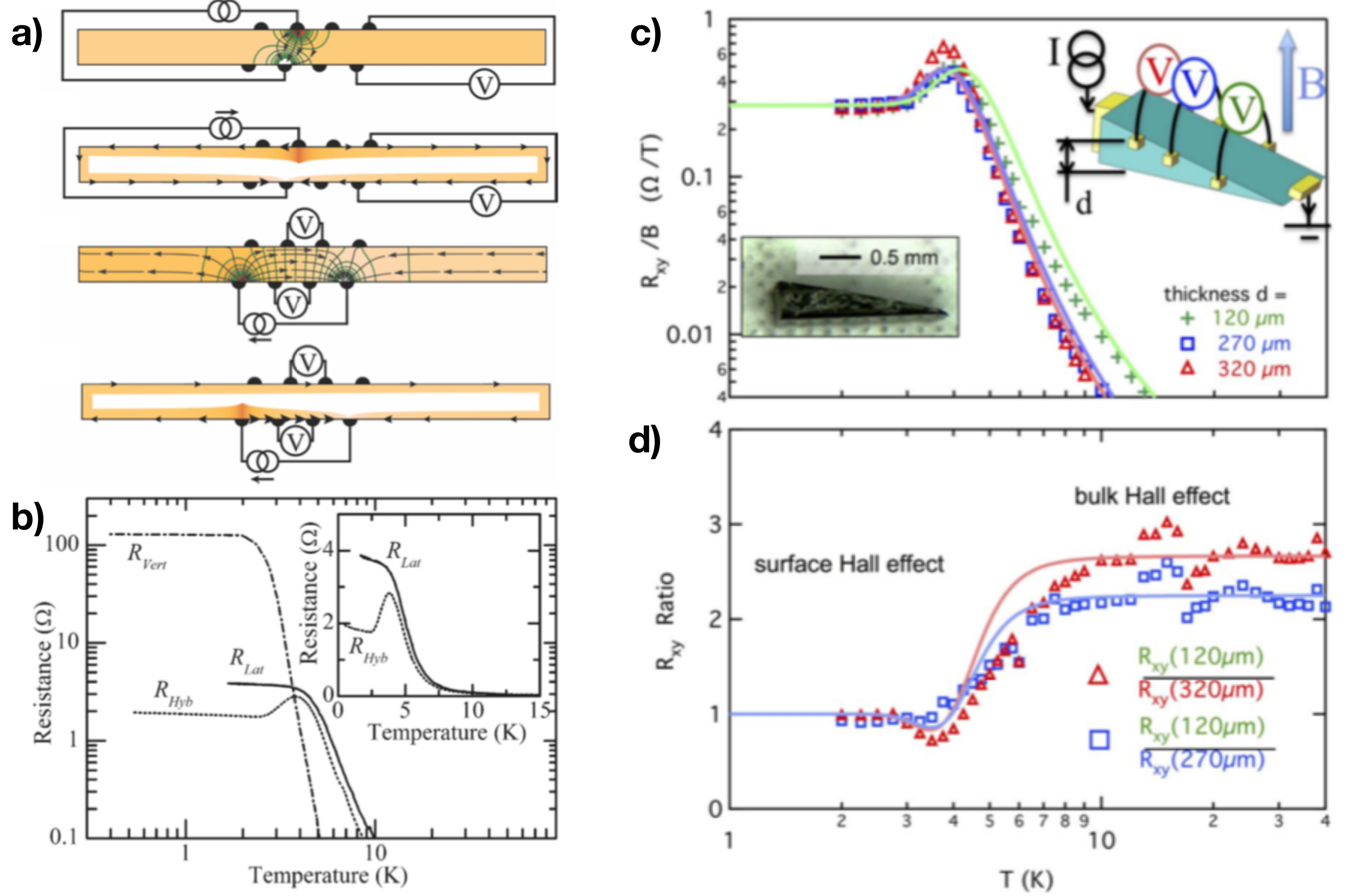}
\caption{(a)~Schematic diagram of the cross section of the sample and electrical contact configurations. Arrows indicate current direction, and lines indicate equipotentials. Reproduced from Wolgast~\textit{et~al.}~\cite{Fis_Wolgast2013}. (b)~Log-log plot of the experimentally determined electrical resistance vs. temperature in different configurations. Inset shows a linear plot of $R_\text{lat}$ and $R_\text{hybrid}$. Reproduced from Wolgast~\textit{et al.}~\cite{Fis_Wolgast2013}. (c)~Temperature dependence of Hall resistance of SmB$_6$ at different thicknesses. Reproduced from Kim~\textit{et al.}~\cite{Fis_Kim2013}. (d)~Hall resistance between different thicknesses. Reproduced from Kim \textit{et al.}~\cite{Fis_Kim2013}.}
\label{fig:Fig2-surface}
\end{figure*}

In 2013, three independent groups reported clever ways of probing whether the resistance saturation is caused by surface states. The first way involves non-local measurements, as shown in Fig.~\ref{fig:Fig2-surface}\,(a)~\cite{Fis_Wolgast2013}. Eight coplanar electrical contacts were attached on the $(100)$ and $(\overline{1}00)$ surfaces of a polished flux-grown SmB$_6$ single crystal. The standard configuration used in previous measurements, $R_\text{lat}$, cannot distinguish surface and bulk conduction; however, measurements using contacts on opposite sides of the samples could. For instance, a vertical measurement, $R_\text{vert}$, is obtained by flowing current from one side of the sample to the opposite side and measuring the voltage drop on another pair of contacts also placed in opposite sides. Further, a hybrid measurement, $R_\text{hybrid}$, is obtained by flowing current on one side of the sample and measuring the voltage on the opposite side. Figure~\ref{fig:Fig2-surface}\,(b) shows the experimental resistance, which is in agreement with the simulations taking into account the low-temperature surface-dominated conduction.

The second way of measuring the surface contribution with dc electrical resistance measurements is by polishing the sample to a well-defined wedge, as shown in the inset of Fig.~\ref{fig:Fig2-surface}\,(c)~\cite{Fis_Kim2013}. By attaching several Hall voltage leads along the length of the sample, one can directly measure the thickness-dependent Hall response, as shown in the main panel of Fig.~\ref{fig:Fig2-surface}\,(c). If dominated by the bulk, the Hall resistance should be inversely proportional to the thickness; however, if conduction occurs mainly on the surface, the Hall resistance should be independent of the thickness. In fact, the Hall resistance ratios between different thicknesses of a flux-grown SmB$_6$ sample become constant below about 4~K, as shown in Fig.~\ref{fig:Fig2-surface}\,(d). The authors are also able to fit the experimental Hall data with a simple two-channel conduction model containing a temperature-independent surface channel and an activated bulk channel.

The third approach makes use of quasiparticle tunneling spectroscopy, which measures the bulk density of states~\cite{Fis_Zhang2013}. The authors argue that, if the resistance saturation below 4~K were due to coherent transport from bulk in-gap states, the electronic structure would change and lead to a zero-bias peak. Their results, however, indicate that the bulk density of states is robust below 10~K, which is taken as evidence for a metallic surface state. Several resistivity studies followed the initial reports, including the use of ionic liquid gating, which supports the scenario of an insulating bulk at high temperatures and metallic surface states that dominate at low temperatures~\cite{Fis_Syers2015}.\index{surface states!in SmB$_6$}\index{SmB$_6$!surface states}

We note that all the measurements discussed so far were performed in flux-grown SmB$_6$. To our knowledge, there are fewer similar experimental investigations on floating-zone samples, and the community would benefit from systematic dc electrical resistivity measurements on these crystals. Here we mention two representative reports. In the first one, the authors performed $R_\text{vert}$ and $R_\text{hybrid}$ measurements on floating-zone samples and conclude that there is not only surface conduction at low temperatures, but also an additional residual bulk conduction possibly arising from a valence-fluctuation induced hopping\index{hopping} from bulk in-gap states \cite{Fis_Gabani2015, Fis_Batkova2018}. The second report performed a systematic resistivity measurement in different cuts of a floating zone grown crystal, and no significant plateau was observed below about 4~K, only a "knee" \cite{Fis_Phelan2014}. Systematic growths in the presence of carbon were then performed because carbon is a common impurity in boron. Remarkably, a thickness-independent plateau emerged in C-doped SmB$_6$ crystals. A sensible possibility suggested by the authors is that both topological and trivial surface states coexist\index{surface states!topologically trivial} as non-topological surface states may arise from chemical variations at the surface. For instance, as we will discuss below, the valence\index{SmB$_6$!mixed valence} of Sm is observed to change to ${3+}$ at the surface (section \ref{Fisk-bulk-neutron}), and the polar [001] surface is prone to polarity-driven surface states \cite{Fis_Zhu2013}.\index{surface states!in SmB$_6$}\index{SmB$_6$!surface states} It is worth noting that a number of alternative theoretical proposals for the surface conductivity of SmB$_6$ have been put forward, including trivial surface states\index{surface states!topologically trivial} discussed above, impurity bands, phonon bound states due to magnetoelastic coupling, Wigner lattice,\index{Wigner crystallization} and Mott minimum conductivity.\index{Mott minimum conductivity} We therefore conclude that, though the community appears to agree that SmB$_6$ hosts surface states, the topological nature of these states remains unsettled.

In spite of the many possible scenarios for its origin, it is fair to state that surface conduction at low temperatures has been established via dc electrical resistivity. As a result, many more questions arise. Why is there a substantial variation in the low-temperature resistivity in Fig.~\ref{fig:Fig1-dc-resistance}? Can one probe the bulk resistance of SmB$_6$ without the influence of the surface states? What is the origin of the feature at $15$~K? And finally, are these surface states topologically protected? The answer to the last question is still disputed, and the rest of this chapter will overview experimental results to help the reader reach a conclusion.

\begin{figure*}[!b]
\includegraphics[width=\columnwidth]{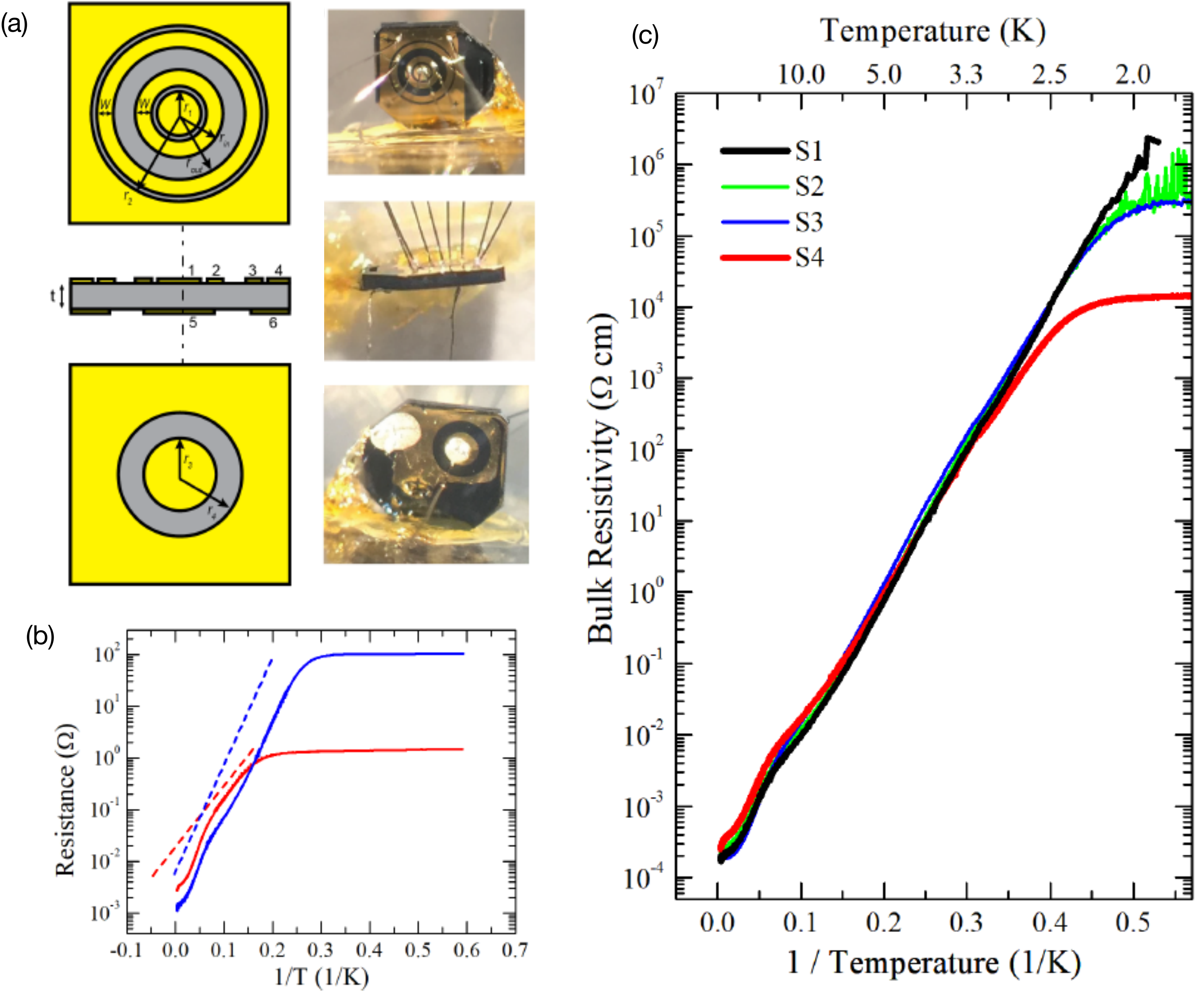}
\caption{(a)~Schematic diagram (left) and photograph (right) of the transport geometry used in double-Corbino measurements. (b)~Electrical resistance as a function of reciprocal temperature for flux-grown SmB$_6$ crystals prepared for Corbino measurements (blue solid curve) and as-grown (red solid curve). (c)~Bulk resistivity as a function of reciprocal temperature for a stoichiometric sample (S1) as well as samples grown with Sm vacancies (S2\,--\,S4). Figures Reproduced from Eo \textit{et al.}~\cite{Fis_Eo2019}.}
\label{fig:Fig3-Corbino}
\end{figure*}

The answer to the first question likely lies on the experimental observations that the bulk-to-surface ratio influences the low-temperature crossover and that small subsurface cracks increase the surface conductivity \cite{Fis_Wolgast2015}. As a result, the surface saturation will depend on how the sample was prepared\,---\,in particular, whether it is as-grown, roughly polished, or finely polished. As the surface quality improves (i.e. no cracks), the conductivity of the surface is reduced. This mechanism further explains the apparent high carrier density obtained in Hall measurements.

The evident consequence of the presence of two conduction channels, one of them with varying conductivity, is that one cannot use the typical inverted resistance ratio (i.e. $\text{IRR} = R_\text{2\,K}/R_\text{300\,K}$) as a good measure of the quality of the sample. This brings us to our next question: can one probe the bulk resistance of SmB$_6$ without the influence of the surface states? The answer is yes, via Corbino-disk measurements \cite{Fis_Eo2019}. The transport geometry for these measurements is shown in Fig.~\ref{fig:Fig3-Corbino}\,(a). The flux-grown samples are first finely polished with a final step of aluminum oxide slurry of 0.3~$\mu$m. The Corbino patterns are then fabricated on the sample via photolithography followed by e-beam evaporation of Ti/Au (20\,\AA/1500\,\AA). By preparing two Corbino disks\index{SmB$_6$!dc electrical conductivity!Corbino method} on opposite surfaces, one can measure either the standard resistivity using just one of the Corbino disks or the so-called ``inverted'' resistivity by applying current on one Corbino and measuring voltage on the opposite one. Figure~\ref{fig:Fig3-Corbino}\,(b) shows the difference in standard resistance between a sample prepared by the Corbino method (blue curve) and an as-grown sample measured by the usual four-probe configuration and no surface preparation (red curve). The significant difference further confirms that the magnitude of the resistance plateau is greatly sensitive to sample preparation. In addition, the extracted gap value also changes from one measurement to another. Figure~\ref{fig:Fig3-Corbino}\,(c) shows the activated plot for the ``inverted" resistivity measurements, which only probe the bulk of the sample. Remarkably, the bulk resistivity of flux-grown SmB$_6$ (S1) rises by 10 orders of magnitude on cooling from 300 to 2~K with no saturation at low temperatures. Samples grown off-stoichiometry with Sm vacancies do show saturation at low temperatures (S2\,--\,S4), indicating that the gap is robust against point defects and strikingly exponential over a range of temperature one would normally expect some deviations due to temperature dependent scattering rate. An open question is whether floating-zone samples exhibit a similar behavior.

The feature in dc electrical resistivity at about 15~K is not often discussed, but it is reproducible. A recent theoretical framework explains this feature by modeling SmB$_6$ as an intrinsic semiconductor with an accumulation length that diverges at cryogenic temperatures \cite{Fis_Rakoski2017}. The self-consistent solution to Poisson's equation taking into account surface charges leads to band bending in the valence and conduction bands as well as to a crossover at about 12~K to bulk-dominated conduction dominated by surface effects. The authors argue that band bending effects explains why the activated gap obtained from dc electrical resistance measurements is smaller than the gap obtained from spectroscopy measurements: spectroscopic methods measure the gap near the surface, whereas transport probes the average gap over the bulk.

We end this section by mentioning dc electrical resistivity measurements on doped or irradiated SmB$_6$. Small amounts ($\sim$3\%) of Gd in flux-grown SmB$_6$ were shown to destroy the surface conduction, whereas the resistance saturation remains intact for samples doped with nonmagnetic Y and Yb at the same concentration level \cite{Fis_Kim2014}. This result was taken as evidence for topological surface states,\index{surface states!in SmB$_6$}\index{SmB$_6$!surface states} which is destroyed by impurities\index{impurities} that break time-reversal symmetry. Similarly, magnetic and nonmagnetic ion irradiation was used to damage the surface layers of flux-grown SmB$_6$ \cite{Fis_Wakeham2015, Fis_Wakeham2016}. The dc resistance results suggest that the surface state is not destroyed by ion irradiation, but instead it is reconstructed below the poorly conducting damaged layer, whether the damage was caused by a magnetic or a nonmagnetic ion. A recent doping study on floating-zone samples has investigated the dc electrical resistivity of SmB$_6$ doped with lanthanum, europium, ytterbium and strontium, revealing a complex response \cite{Fis_Gabani2016}.
\index{SmB$_6$!dc electrical conductivity|)}

\subsection{Tunneling spectroscopy and thermopower}\index{SmB$_6$!tunneling spectroscopy|(}\index{SmB$_6$!thermopower|(}
\label{Fisk-surface-stm}

\begin{figure*}[t]
\centerline{\includegraphics[width=1.02\textwidth]{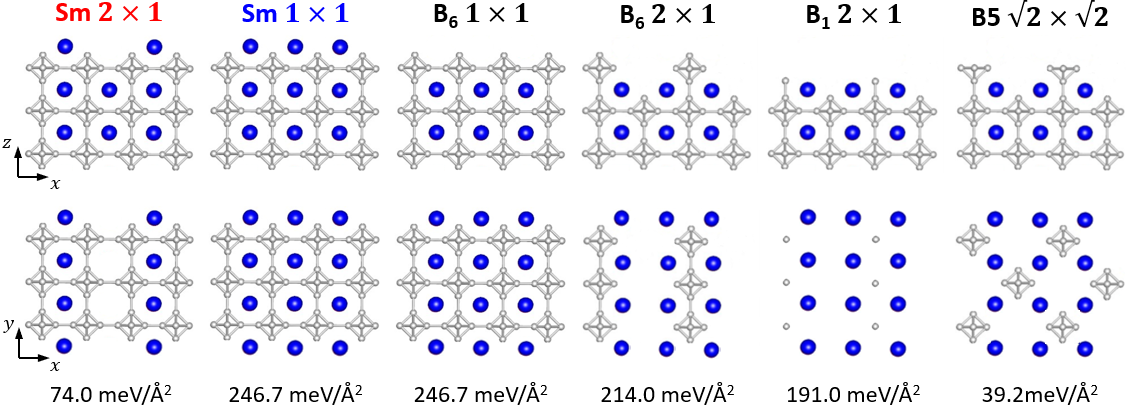}}
\caption{Surface formation energy of six different surface reconstructions\index{surface reconstruction} in SmB$_6$.
Reproduced from Matt \textit{et al.} \cite{Fis_MattPirie20}.}\index{SmB$_6$!surface energy}
\label{fig:Fig4-SurfaceRec}
\end{figure*}

The measurements presented in the previous section are an average over the whole surface of SmB$_6$, whereas scanning tunneling microscopy and spectroscopy measurements are able to probe the surface of SmB$_6$ on an atomic level. SmB$_6$, however, lacks a natural cleavage plane. As a result, various surface terminations are possible, and surface domains might emerge. Independent groups have performed density functional theory calculations of the surface formation energy for different surface terminations, as exemplified in Fig.~\ref{fig:Fig4-SurfaceRec} \cite{Fis_Kim2014a, Fis_Sun2018, Fis_MattPirie20}. The lower energy pentaboride termination is argued to form disordered regions \cite{Fis_MattPirie20}, whereas the \mbox{$2\!\times\!1$} Sm termination was argued to be ordered and non-polar \cite{Fis_PirieLiu20, Fis_Rossler2014}, though this assignment has been questioned recently \cite{Fis_Herrmann2018}.

\begin{figure*}[!ht]
\includegraphics[width=\textwidth]{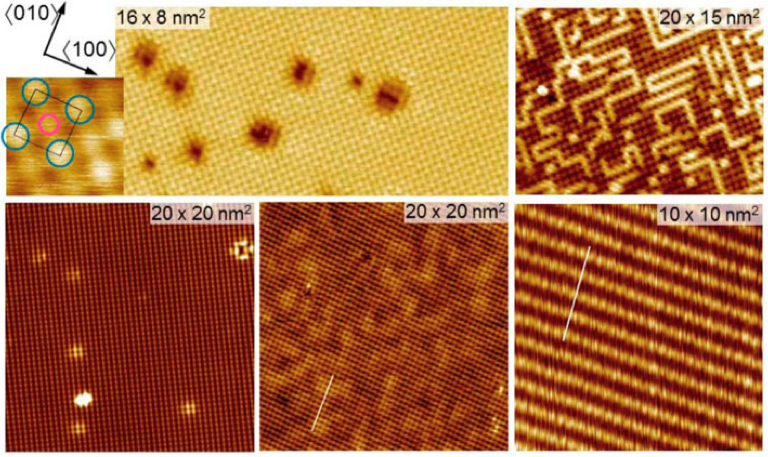}
\caption{Summary of representative topographies observed on cleaving a (001) SmB$_6$ surface. Reproduced from R\"{o}{\ss}ler \textit{et al.} \cite{Fis_Rossler2016}. The upper left and center topographies display Sm-terminated surfaces, whereas the bottom left and center topographies display B-terminated surfaces. The upper right topography shows a reconstructed disordered surface, whereas the bottom right topography shows a \mbox{$2\!\times\!1$} reconstruction.}\index{SmB$_6$!surface topography}\index{scanning tunneling microscopy}
\label{fig:Fig5-STM}
\end{figure*}

In fact, reports on flux-grown SmB$_6$ observed several distinct surface terminations on the same surface \cite{Fis_Ruan2014, Fis_Rossler2014, Fis_Jiao2016, Fis_PirieLiu20, Fis_Sun2018}. The crystals are typically cleaved \textit{in~situ}, and four main categories of 001 surface terminations have been observed: (i)~Sm-terminated surfaces; (ii)~B-terminated surfaces; (iii)~disordered reconstructed surfaces; and (iv)~ordered reconstructed surfaces (e.g. \mbox{$2\!\times\!1$}), as shown in Fig.~\ref{fig:Fig5-STM}. Further, the temperature at which the crystals are cleaved appear to play a significant role in determining the surface termination \cite{Fis_Zabolotnyy2018}. At room temperature, Ruan~\textit{et al.} cleaved the samples through the B$_6$ octahedra, exposing a donut-shaped structure~\cite{Fis_Ruan2014}, whereas the cleavage planes appear between octahedra at lower temperatures~\cite{Fis_Rossler2014}. Experimental reports on floating-zone samples, however, argued that only one topography is observed on a given surface, though the authors recognize they did not scan the entirety of the cleaved area \cite{Fis_Herrmann2018}. For details on the surface termination assignment, we refer the reader to~\cite{Fis_Rossler2016}.

We recall that the bulk-truncated (001) surface of SmB$_6$ is polar, which could give rise to band bending, charge puddles, and the surface reconstructions\index{surface reconstruction} mentioned above. Further, surface reconstructions could generate metallic surface layers, making the observation of a topologically non-trivial surface state challenging. In spite of the outstanding issues with surface termination, there are similarities between different reports we would like to highlight. First, a decrease in the ${\rm d}I/{\rm d}V$ spectrum is observed in scanning tunneling microscopy experiments at about 10--20~meV at temperatures below $\sim$35~K and attributed to the opening of the hybridization gap. The conductance, however, does not vanish at zero voltage, indicating a finite density of states at $E_{\rm F}$, which is consistent with both an incomplete gap and an additional conductance channel at the surface. At lower temperatures, several features emerge within the gap. In particular, a large sharp peak is observed at around $-7$~meV, even in reconstructed surfaces, though the origin of such feature is not agreed on. Jiao~\textit{et al.} argue that this feature contains not only a bulk contribution, but also a surface component below $7$~K \cite{Fis_Jiao2016}. In Gd-doped SmB$_6$, this $-7$~meV feature is destroyed at the impurity site with a healing length of approximately 1~nm in the vicinity of a defect. In addition, a magnetic tip is found to have significant effect on the local electronic structure. These findings are consistent with the expectation that the protected nature of a topological surface conducting state can be destroyed by time-reversal symmetry breaking \cite{Fis_Jiao2018}. Evidence for topologically non-trivial surface states\index{surface states!in SmB$_6$}\index{SmB$_6$!surface states} is also presented by quasiparticle interference spectroscopy \cite{Fis_PirieLiu20}, which images the formation of linearly dispersing surface states\index{surface states!in SmB$_6$}\index{SmB$_6$!surface states} with heavy effective masses (e.g. $m^{*}\sim 400 m_\text{e}$ at the $X$ point). Thermopower and Nernst effect measurements on the (110) plane of SmB$_6$ also indicate that the metallic surface state has a large effective mass \cite{Fis_Luo2015}. Scanning tunneling spectroscopy measurements and analysis by Herrmann~\textit{et al.} \cite{Fis_Herrmann2018}, however, argue for a modification of the low-energy electronic structure at the surface, which would in turn give rise to topologically trivial surface conductivity\index{surface states!topologically trivial}\index{SmB$_6$!surface conductivity} due to the termination-dependent $4f$-like intensity near the Fermi level.

Planar tunneling spectroscopy measurements are also sensitive to the surface, and a Pb-SmB$_6$ junction has been used to probe the spectroscopic properties of SmB$_6$ \cite{Fis_Park2016, Fis_Sun2017}. The differential conductance on both (100) and (110) surfaces display a peak at about $-21$~mV attributed to the bulk hybridization gap, in agreement with scanning tunneling spectroscopy measurements discussed above. Below about 15~K, the conductance increases, taken as an indication of a stronger surface state contribution, in agreement with the feature in dc resistivity discussed above. A V-shaped linear conductance is then observed at low bias arguably to the expected Dirac fermion density of states. This linearity, however, ends at about 4~mV and is attributed to inelastic tunneling via emission and absorption of bosonic excitations, i.e., spin excitons\index{spin exciton} on the surface. The authors conclude, enlightened by theory, that the topological protection of the surface states\index{surface states!in SmB$_6$}\index{SmB$_6$!surface states} is incomplete owing to their strong interaction with bulk spin excitons.\index{spin exciton} As a result, low-energy protected surface states may only exist below 5\,--\,6~K when the interaction with spin excitons become negligible.

Finally, we note that no quantum oscillations\index{SmB$_6$!quantum oscillations} were observed in dc resistance, planar tunneling spectroscopy or thermopower measurements described so far, though torque magnetometry measurements initially revealed oscillatory patterns as we will describe in the next section.
\index{SmB$_6$!thermopower|)}\index{SmB$_6$!tunneling spectroscopy|)}

\subsection{Quantum oscillations}
\begin{figure*}[!b]
\centerline{\includegraphics[width=1.02\columnwidth]{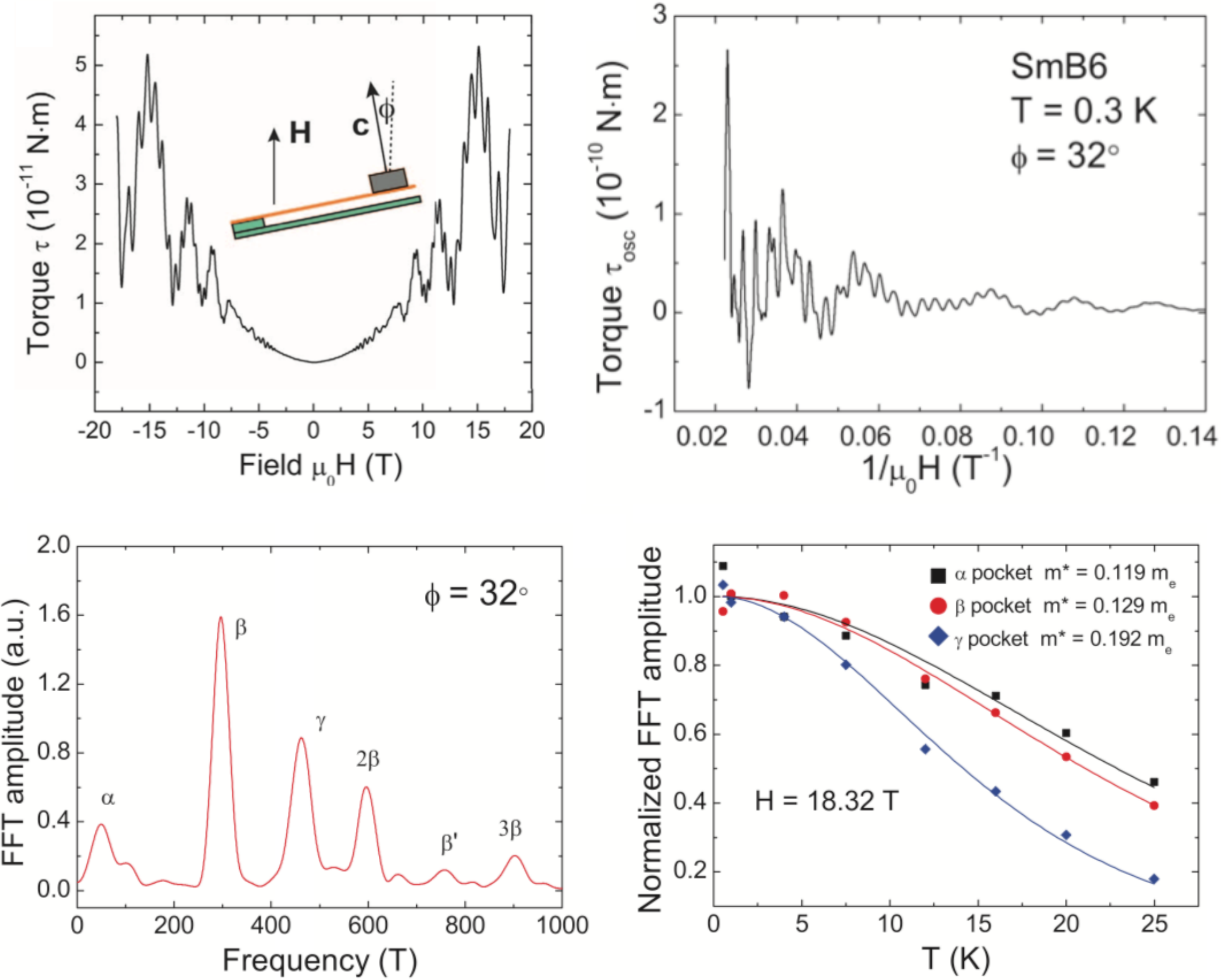}}
\caption{(a)~Magnetic torque of flux-grown SmB$_6$ at 300~mK as a function of magnetic field. (b)~Oscillatory torque as a function of reciprocal field. (c)~Fast Fourier Transform (FFT) of the oscillatory torque. (d)~Temperature dependence of the FFT amplitude. Reproduced from Li \textit{et al.} \cite{Fis_Li2014}.}
\label{fig:Fig6-QO}
\end{figure*}

\index{SmB$_6$!quantum oscillations}
\label{Fisk-surface-qo}
The first report of quantum oscillations in SmB$_6$ used flux-grown crystals \cite{Fis_Li2014}. Li~\textit{et al.} used torque magnetometry to map the oscillation pattern as a function of temperature and angle, as exemplified in Fig~\ref{fig:Fig6-QO}\,(a,b).

Three pockets were observed at $\alpha = 29$~T, $\beta = 286$~T and $\gamma = 385$~T with unexpectedly light effective masses of $m/m_\text{e} = 1.1, 0.8$ and $0.36$, respectively [Fig~\ref{fig:Fig6-QO}\,(c,d)]. The presence of light quasiparticles was argued theoretically to be due to a reduction of the Kondo effect on the surface associated with Kondo breakdown~\cite{Fis_Erten2016}.\index{Kondo breakdown} Both the effective masses and the angular dependence of the observed quantum oscillations, however, resemble that of aluminum, the flux used to prepare the single crystalline samples used in the experiment. Li~\textit{et al.} argued that the aluminum pellets used in the growth are polycrystalline and, as a result, could not generate the observed angular dependence. The authors therefore concluded that the measured Fermi surface cross sections scaled as the inverse cosine function of the magnetic field tilt angles, demonstrating the two-dimensional nature of the conducting (surface) states. The same group later showed that, though the angular dependence of the frequency of the $\beta$ branch is fourfold symmetric, the angular dependence of the amplitude of the same branch is twofold symmetric. This result was taken as evidence of multiple light-mass surface states in SmB$_6$ \cite{Fis_Xiang2017}.\index{surface states!in SmB$_6$}\index{SmB$_6$!surface states}

\begin{figure*}[!ht]
\includegraphics[width=\columnwidth]{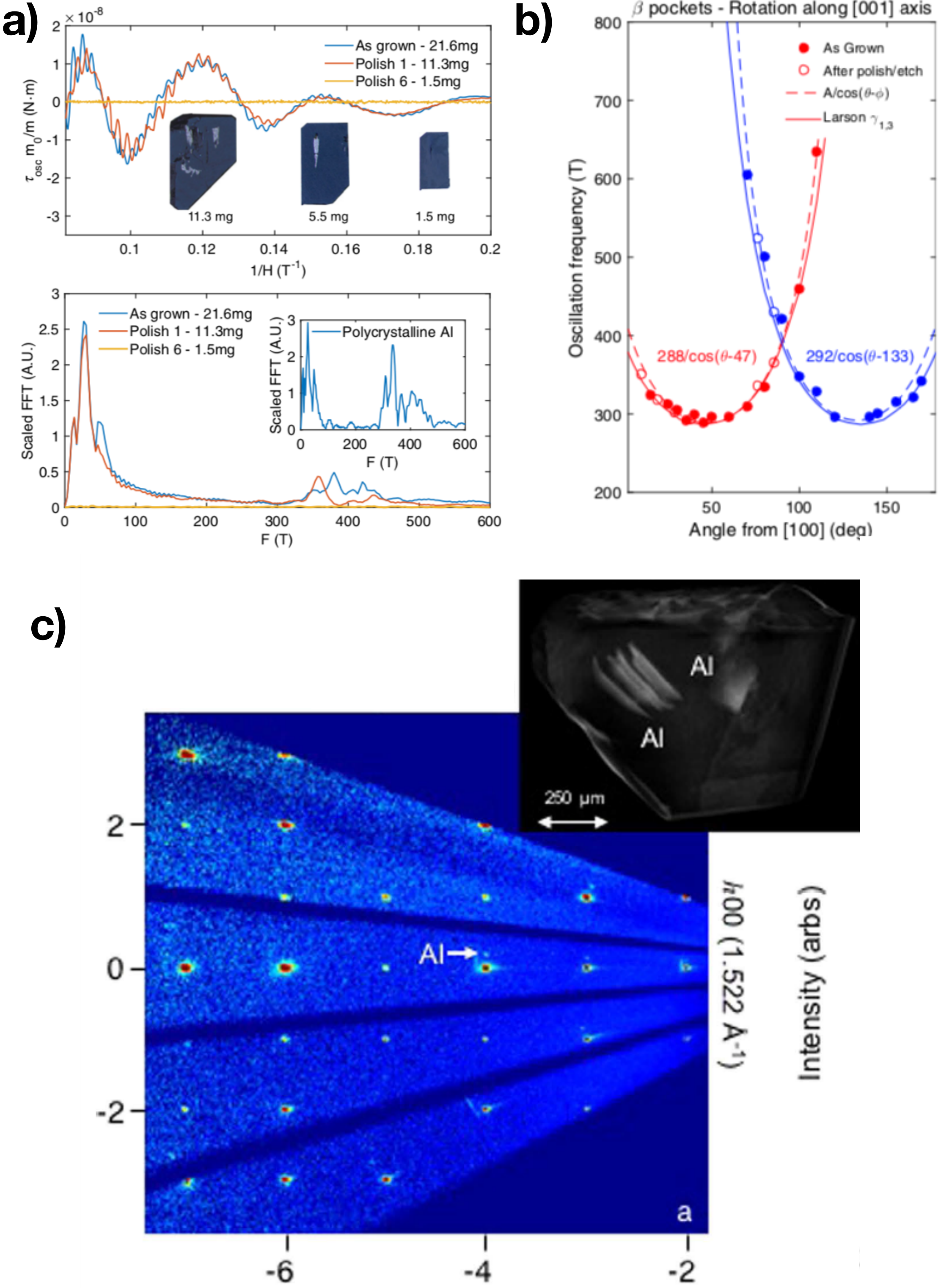}\vspace{-3pt}
\caption{\hspace{-2pt}(a)~Thickness dependence of the oscillatory torque\index{oscillatory torque} in a flux-grown SmB$_6$ crystal. (b)~FFT of the oscillatory torque. The FFT was scaled by the sample mass. Inset shows the FFT of polycrystalline Al. Reproduced from Thomas~\textit{et al.} \cite{Fis_Thomas2019}. (b)~Angular dependence of the oscillatory frequency compared with the $1/\cos\theta$ behavior as well as with the early report on Al by Larson and Gordon \cite{Fis_Larson1967}. (c)~Diffraction pattern of a flux-grown SmB$_6$ crystal and an x-ray computed tomography image showing the presence of aluminum inclusions. Reproduced from Phelan~\textit{et al.}~\cite{Fis_Phelan2016}.\vspace{-3em}}
\label{fig:Fig7-QO}
\end{figure*}

A recent report on this subject has revisited the quantum oscillation patterns of flux-grown SmB$_6$ \cite{Fis_Thomas2019}. Thomas~\textit{et al.} were only able to observe quantum oscillations in about 50\% of the samples, and these samples tended to have larger thickness. As mentioned in the section \ref{Fisk-growth}, the main disadvantage of the flux technique in this case is that flux could become an inclusion. As a result, a thickness-dependent study was performed on the thicker samples displaying quantum oscillations. Care was taken to polish only one side of the sample to maintain any quantum oscillations coming from the other surfaces intact. After polishing the samples, large unconnected aluminum deposits were revealed, as shown in Fig.~\ref{fig:Fig7-QO}\,(a), and these deposits could be easily removed with dilute hydrochloric acid. After removing all aluminum deposits, the thickness of the samples is about 100\,--\,200 microns, and no quantum oscillations\index{SmB$_6$!quantum oscillations} were observed to 45~T.

This experiment demonstrates that quantum oscillations in flux-grown SmB$_6$ originates from flux inclusions; however, the original aluminum pellet used in the growth is polycrystalline. To solve this apparent contradiction, one needs to take into account the fact that aluminum also crystallizes in a cubic structure with a lattice parameter that is only 2\% smaller than that of SmB$_6$. In fact, x-ray tomography performed previously by a third group showed the presence of several aluminum deposits co-crystallizing with the SmB$_6$ host in typical millimeter-sized samples Fig.~\ref{fig:Fig7-QO}\,(a) \cite{Fis_Phelan2016}. Precisely because several aluminum deposits may exist within one large SmB$_6$ single crystal, the amplitude of the quantum oscillation pattern may not be C$_4$ symmetric if the deposits are not perfectly aligned. In section \ref{Fisk-bulk-qo}, we will discuss bulk quantum oscillation results in floating-zone samples.

\subsection{Angle-resolved photoelectron spectroscopy~}\index{SmB$_6$!photoelectron spectroscopy}
\label{Fisk-surface-arpes}

A detailed overview of angle-resolved photoemission spectroscopy measurements on SmB$_6$ is beyond the scope of this book chapter. In principle, spin- and angle-resolved photoemission spectroscopy would be one of the most direct ways of probing the surface state dispersion and spin texture; however, the issues with surface terminations mentioned above as well as the small hybridization gap of SmB$_6$ hinder a consensus.

As in the case of scanning tunneling microscopy, there are reports in favor of either topologically protected surface states \cite{Fis_Neupane2013, Fis_Jiang2013, Fis_Xu2014, Fis_Ohtsubo2019} or trivial surface conductivity\index{surface states!topologically trivial} \cite{Fis_Zhu2013, Fis_Hlawenka2018}. For recent reviews of the subject, we invite the reader to refer to \cite{Fis_Denlinger2014, Fis_Allen2016, Fis_MattPirie20, Fis_Herrmann2018}.

\subsection{Thin films and nanowires}\index{SmB$_6$!thin films|(}\index{SmB$_6$!nanowires|(}
\label{Fisk-surface-films}

Though this chapter focuses on bulk and surface properties of SmB$_6$ single crystals, in this section we give a brief overview of recent efforts into the synthesis and characterization of SmB$_6$ thin films and nanowires. Early reports on thin film synthesis can be found in Refs. \cite{Fis_Batko1990, Fis_Waldhauser1998}. In 2014, polycrystalline SmB$_6$ thin films were synthesized by co-sputtering of SmB$_6$ and boron target within a combinatorial composition-spread approach \cite{Fis_Yong2014}. A similar approach was used in 2017 by Petrushevsky~\textit{et al.} \cite{Fis_ShavivPetrushevsky2017}. Growth attempts using pulsed laser deposition, molecular beam epitaxy or sputtering using a single target often result in thin films with substantial boron deficiency \cite{Fis_Yong2014, Fis_Shishido2015, Fis_Cen2017}. Batkova \textit{et al.}, however, have recently reported the preparation of stoichiometric SmB$_6$ thin films via pulsed laser deposition \cite{Fis_Batkova2018a}.

Electrical resistance measurements on SmB$_6$ thin films typically display a semiconducting behavior with a resistance ratio, $R_\text{2\,K}/R_\text{300~K}$ much smaller than that of the bulk, in agreement with the larger surface-to-bulk ratio \cite{Fis_Yong2014, Fis_Batkova2018a}. Initial point-contact spectroscopy\index{SmB$_6$!point-contact spectroscopy} using a superconducting tip revealed the presence of Andreev reflection,\index{Andreev reflection} suggesting the presence of surface states and of a transparent SmB$_6$/superconductor interface \cite{Fis_Yong2014}. Further evidence for a transparent interface was obtained by \textit{in situ} deposition of superconducting Nb layers \cite{Fis_Lee2016}. On one hand, magnetotransport as well as penetration depth measurements on SmB$_6$ thin films were argued to be consistent with topological surface states \cite{Fis_Yong2015, Fis_Liu2018}. On the other hand, Li \textit{et al.} observe that the electrical resistivity of SmB$_6$ thin films is thickness dependent, in apparent contradiction with the surface conductivity scenario \cite{Fis_Li2018}. Li~\textit{et al.} also show that SmB$_6$ thin films display large spin-orbit torque. Nevertheless, a recent point-contact spectroscopy measurement reports the observation of perfect Andreev reflection\index{Andreev reflection} in a heterostructure formed by insulating SmB$_6$ and superconducting YB$_6$. This observation was understood as a manifestation of Klein tunneling due to the proximity-induced superconducting state in a topological insulator \cite{Fis_Lee2019}.

Rare-earth hexaboride nanowires have been synthesized by a variety of methods including vapor-liquid-solid mechanism \cite{Fis_Brewer2007, Fis_Xu2013}, chemical vapor deposition with \cite{Fis_Zhang2005} and without \cite{Fis_Xu2008} a catalyst, and palladium-nanoparticle-assisted chemical vapor deposition \cite{Fis_Brewer2011}. Electrical resistance measurements are argued to be consistent with the presence of topological surface states \cite{Fis_Zhou2016, Fis_Han2016, Fis_Lin2017, Fis_Gan2019}.

\vspace{-2pt}\section{Bulk properties}\vspace{-1pt}
\label{Fisk-bulk-section}

\subsection{ac electrical conductivity}\index{SmB$_6$!ac electrical conductivity|(}
\label{Fisk-bulk-ac}

As discussed in section \ref{Fisk-surface-dc}, the bulk of flux-grown SmB$_6$ has been shown to be remarkably insulating, with a 10-order-of-magnitude increase in dc resistivity on cooling from room temperature to 2~K. The bulk ac conductivity of SmB$_6$, however, is several orders of magnitude larger than that of any known impurity band, and this is the next puzzle we would like to address.

Because of the large index of refraction of hexaborides and the particularly large ac conduction in SmB$_6$, previous optical measurements were typically performed in reflection mode, relying on the Kramers-Kronig transformation \cite{Fis_Travaglini1984, Fis_Jackson1984, Fis_Kimura1994, Fis_Nanba1993, Fis_Tytarenko2016}. Further, temperature- and thickness-dependent transmission experiments were missing until the renewed interest in SmB$_6$. Early low-temperature ac conductivity measurements on floating-zone SmB$_6$ in the range from 0.6 to 4.5~meV provided evidence for a 19~meV energy gap and an additional narrow donor-type band at 3~meV below the conduction band \cite{Fis_Gorshunov1999}. The properties of SmB$_6$ below 8~K were first attributed to localized carriers in the narrow band responsible for hopping\index{hopping} conductivity \cite{Fis_Gorshunov1999} and later to exciton-polaron complexes \cite{Fis_Sluchanko2000}.

\begin{figure*}[!t]
\includegraphics[width=\columnwidth]{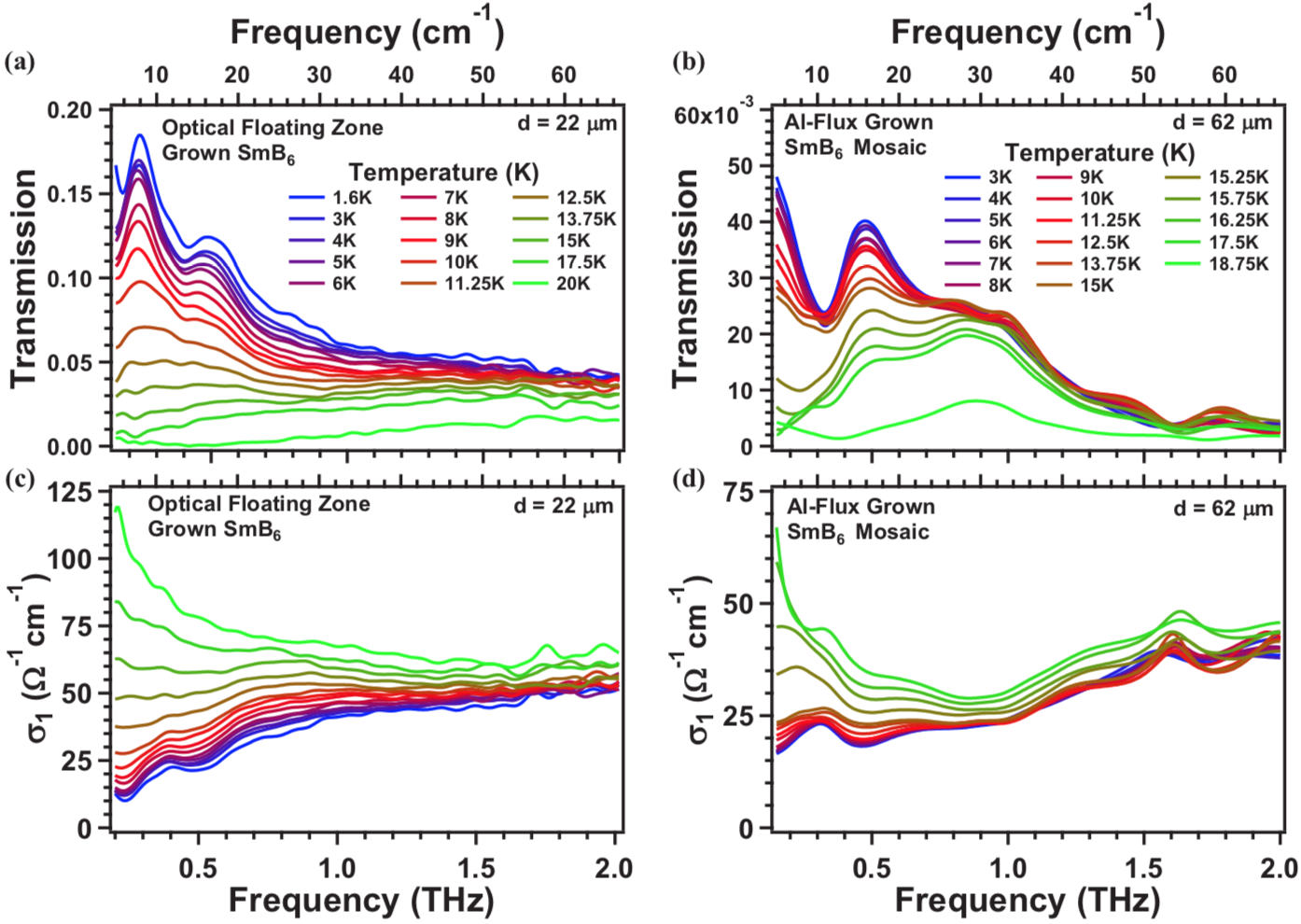}
\caption{Frequency dependence of the complex transmission for (a)~floating zone and (b)~flux-grown SmB$_6$ single crystals with thickness of 22 and $62$~$\mu$m, respectively. Frequency dependence of the real part of the optical conductivity for (c)~floating zone and (d)~flux-grown SmB$_6$ single crystals. Reproduced from Laurita \textit{et al.} \cite{Fis_Laurita2016}.}
\label{fig:Fig8-ac}
\end{figure*}

High-resolution optical measurements in the terahertz frequency range were recently performed to revisit these results. Laurita~\textit{et al.} investigated both floating zone and flux-grown samples with thicknesses smaller than 100 microns \cite{Fis_Laurita2016}. The experimental energy range, $\hbar\omega\approx 1$\,--\,8~meV, is smaller than the spectroscopic gap energy, which allows this technique to probe states within the bulk gap.

Figure \ref{fig:Fig8-ac} shows the frequency dependence of the complex transmission at various temperatures for floating-zone (panel a) and flux-grown (panel b) SmB$_6$ crystals, respectively. It is worth noting that flux-grown samples are smaller, and the required mosaic of samples displays worse signal to noise. Both real and imaginary parts of the complex optical conductivity are therefore extracted from the complex transmission, and the real part of the optical conductivity is shown in Fig.~\ref{fig:Fig8-ac}\,(c,d). At high temperatures, the optical conductivity displays a Drude-like response, i.e., $\sigma_1$ increases with decreasing frequency, indicating the presence of free charge carriers at $E_\text{F}$. Above 30~K, the sample becomes opaque in the THz range. The magnitude of the Drude-like response decreases with decreasing temperature, in agreement with the opening of the hybridization gap. Below 13~K, the ac conductivity increases linearly with frequency before saturating at about 1~THz. This additional conduction channel is termed ``localized'' because it does not contribute to dc transport.

The pressing question here is whether this in-gap conduction is due to impurity states or to exotic neutral excitations,\index{charge-neutral quasiparticles|(}\index{Fermi surface!charge neutral|(} and answering this question is challenging. A recent analysis starts with the conjecture that the localized conductivity response is independent of temperature based on the weak temperature dependence of the conductivity at high frequencies \cite{Fis_Laurita2018}. The conductivity can be in turn modeled as a sum of the localized contribution, the Drude response, and a frequency-independent background. The latter term has been interpreted in previous measurements as a Mott minimum conductivity~\cite{Fis_Gorshunov1999}.\index{Mott minimum conductivity} The dc conductivity can be extracted from the Drude response, providing an activated gap of 4.1~meV, in agreement with dc measurements discussed in section \ref{Fisk-surface-dc}. The frequency independent conductivity is finite only above 12~K and reaches $9.4~\Omega^{-1}$ at 17.5~K.

The localized contribution within the gap dominates the conductivity response at 1.6~K. At about 1~THz (4~meV), the frequency dependence of the conductivity displays a crossover from $\sigma_1 \propto \omega^{0.8}$ to a linear dependence with frequency. The frequency dependence below the indirect gap is inconsistent with the weak Drude peak predicted by disordered Kondo insulator models, which take into account effects of substitution on the $f$-electron site above a percolation threshold \cite{Fis_Schlottmann1992, Fis_Riseborough2003}. A sensible scenario, however, is that SmB$_6$ is below the percolation threshold, in which impurity states are localized. In addition, substitutional (point-defect) impurities\index{impurities} might not be the only source of defects in SmB$_6$. A quantitative comparison between SmB$_6$ and  localization driven insulators (e.g. Si:P at 40\% doping) reveals that the in-gap ac conductivity of SmB$_6$ is about four orders of magnitude larger than the typical impurity band conduction, being comparable to completely amorphous alloys \cite{Fis_Helgren2004, Fis_Geballe2013}. This possibility brings up again the lack of a deep understanding of disorder in this material. In fact, a recent theoretical effort revisits the donor impurity band mechanism in SmB$_6$ taking into account its peculiar "mexican-hat-like" band structure  and shows that the resulting impurity band is in many ways distinct from the conventional semiconductor case \cite{Fis_Skinner2019}. In particular, the critical doping concentration necessary to drive an insulator-metal transition is much larger than in the conventional case. Estimates of the ac conductivity in this framework are in agreement with the experimental results discussed above and provide an explanation for why SmB$_6$ is a robust dc insulator, but an ac conductor. The second sensible scenario is the presence of charge-neutral quasiparticles within the Kondo gap, which could couple to an ac electric field \cite{Fis_Erten2017}. Though this scenario has been proposed theoretically, a qualitative prediction that matches the experimental power law behavior with frequency is missing. This scenario has been also suggested by quantum oscillation and thermal conductivity measurements on floating zone samples, as we will discuss in the next two sections. Finally, we note that the spectral weight of the in-gap conductivity provides further information on the density of charge carriers and effective mass $via$ the conductivity sum rule relation. A rough estimate using the bare mass of the electron gives a charge density of about 1~electron per 1000 unit cells, which is roughly consistent with both an impurity contribution as well as a neutral Fermi surface. Other experimental probes are therefore required to answer this question.

\begin{figure*}[!b]
\centerline{\includegraphics[width=1.01\columnwidth]{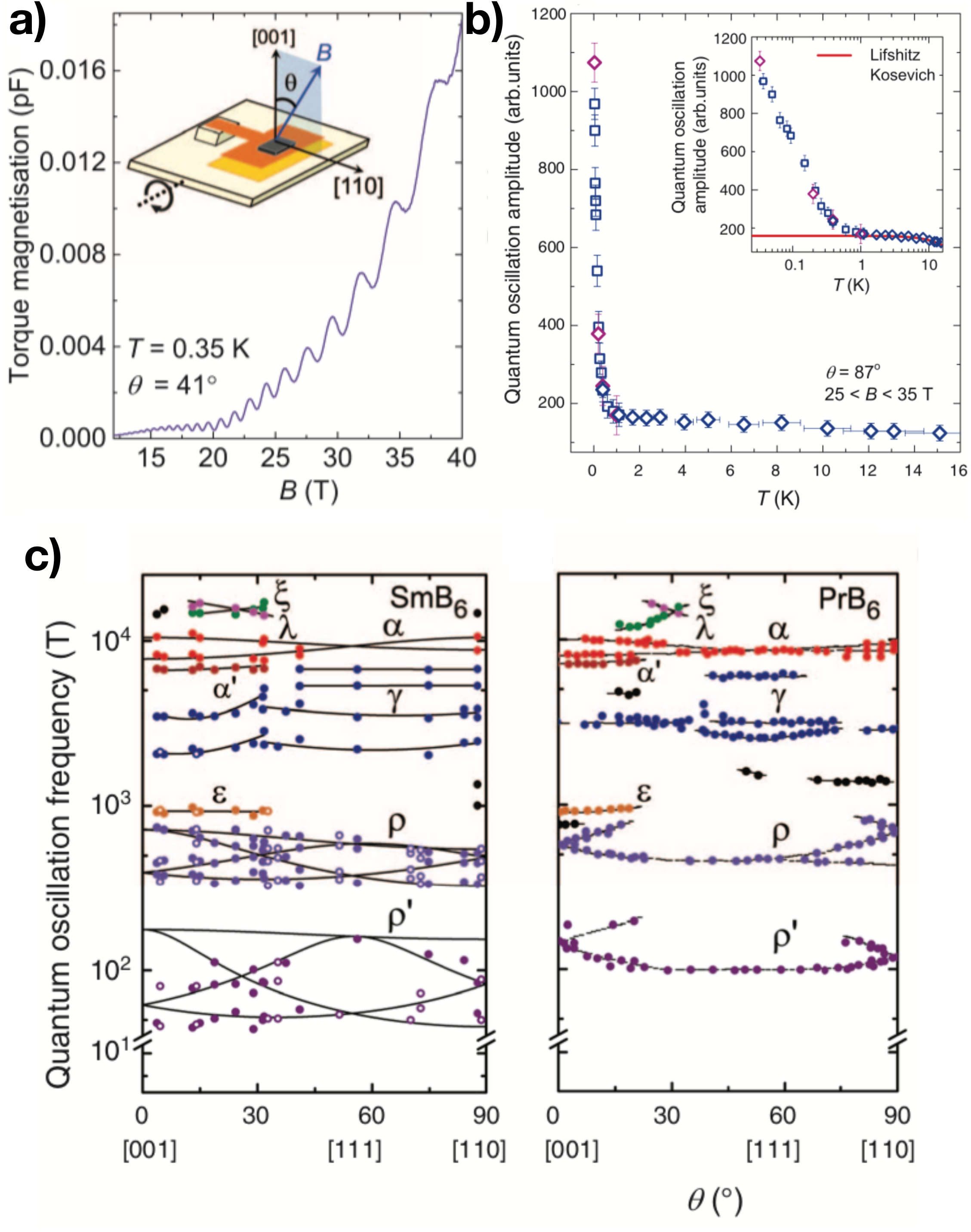}}\vspace{-2pt}
\caption{(a)~Torque signal of a floating-zone SmB$_6$ crystal as a function of magnetic field. (b)~Quantum oscillation amplitude as a function of temperature. (c)~Angular dependence of the quantum oscillation frequency for SmB$_6$ and PrB$_6$. Reproduced from Tan~\textit{et~al.}~\cite{Fis_Tan2015}.\vspace{-3pt}}
\label{fig:Fig9-QO-FZ}
\end{figure*}

\subsection{Quantum oscillations}\index{SmB$_6$!quantum oscillations}
\label{Fisk-bulk-qo}

Shortly after the report of quantum oscillations in flux-grown SmB$_6$, which we discussed in section \ref{Fisk-surface-qo}, similar torque magnetometry experiments were performed on floating-zone crystals \cite{Fis_Tan2015}. The first key remark is that the oscillatory behavior shown in Fig.~\ref{fig:Fig9-QO-FZ}\,(a) is \textit{not} observed in every floating-zone sample. This is an outstanding question which calls for immediate attention, especially considering that flux-grown samples free of aluminum do not show quantum oscillations and that floating-zone rods host compositional changes that are not fully understood.

When quantum oscillations are present in floating-zone samples, they are observed at high frequencies and are independent of the surface details. Further, the angular dependence of the oscillations resembles those of LaB$_6$ and PrB$_6$, as shown in Fig.~\ref{fig:Fig9-QO-FZ}\,(c). Taken together, these observations are inconsistent with surface-driven quantum oscillations and suggest a bulk origin.

Quantum oscillations in floating-zone samples were termed ``unconventional'' owing to the temperature dependence of the quantum oscillation amplitude, shown in Fig.~\ref{fig:Fig9-QO-FZ}\,(b). Above 1~K, the amplitude follows the conventional Lifshitz-Kosevich formula \cite{Fis_Shoenberg1984} with a small effective mass (0.1-0.8~$m_\text{e}$); however, below 1~K the amplitude increases greatly down to base temperature. Importantly, oscillations in the dc electrical resistivity are not observed.

As in the case of ac conductivity measurements, we are again faced with (at least) two possibilities. The first one is that quantum oscillations may arise from spatially unconnected metallic patches from a secondary unknown phase, similar to the aluminum inclusions observed in flux-grown samples. Though Tan~\textit{et al.} state that this possibility appears to be unlikely, it cannot be ruled out at this point. The second possibility is the presence of low-energy neutral excitations within the charge gap of SmB$_6$. Hartstein~\textit{et al.} find that the density of states obtained from their quantum oscillation Fermi surface matches that of specific heat measurements ($\gamma \sim 4(1)$~mJ/mol.K$^{2}$) \cite{Fis_Hartstein2018}. The steep increase in the low-temperature quantum oscillation amplitude also resembles the increase in the specific heat. Further, the low-temperature entropy obtained from the oscillatory pattern remains finite below 1~K, indicating a finite density of states. These observations, combined with thermal conductivity measurements to be discussed in the next section, were taken as evidence for bulk itinerant low-energy excitations that couple to magnetic fields, but not weak dc electric fields.

\begin{figure*}[t]\vspace{-2pt}
\centerline{\includegraphics[width=1.01\textwidth]{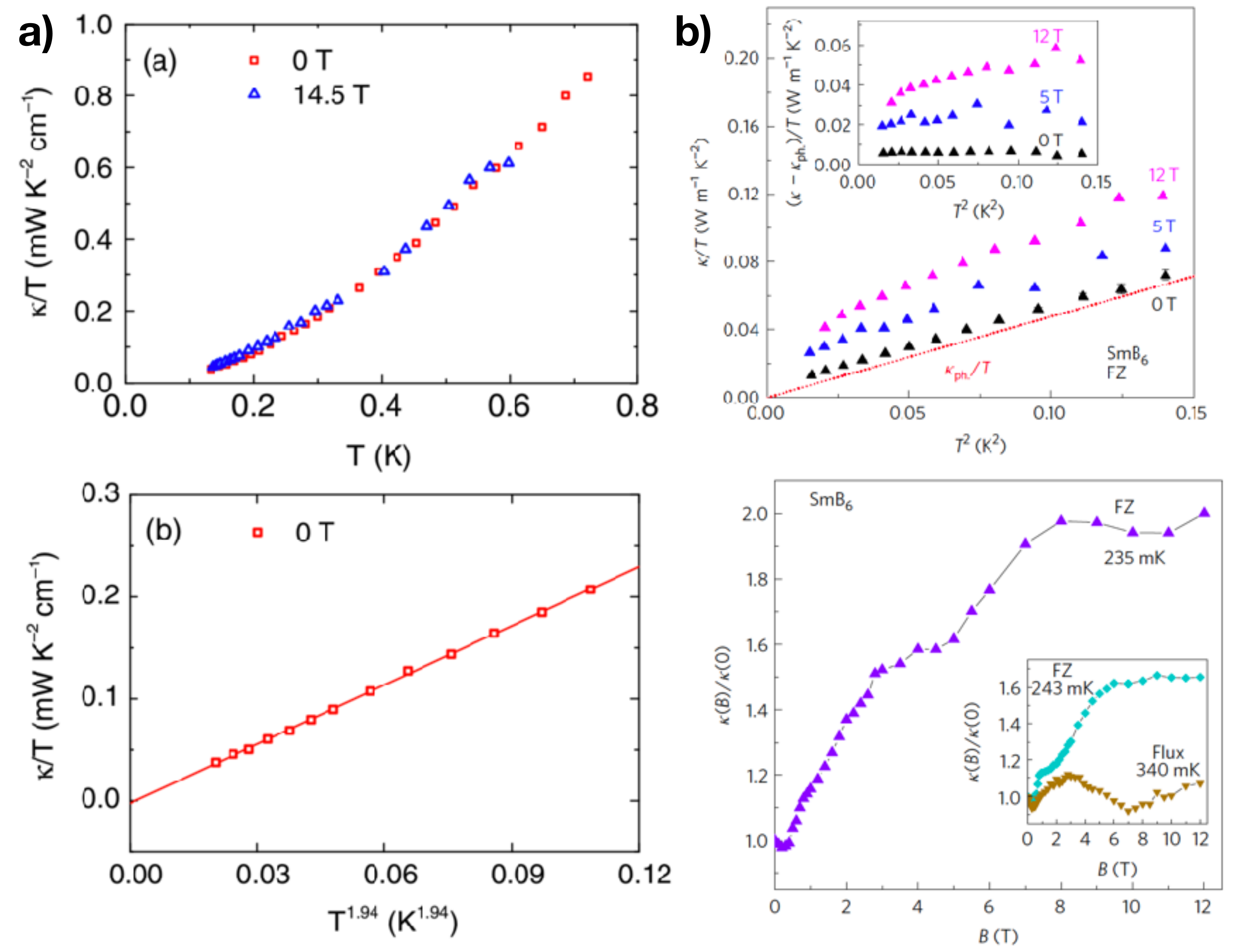}}\vspace{-2pt}
\caption{(a)~Thermal conductivity of a flux-grown SmB$_6$ sample vs. temperature (top) and $T^{1.94}$ (bottom). Reproduced from Xu \textit{et al.}~\cite{Fis_Xu2016}. (b)~Thermal conductivity of a floating-zone SmB$_6$ sample vs. $T^{2}$ (top) and magnetic field (bottom). Reproduced from Hartstein \textit{et al.} \cite{Fis_Hartstein2018}.}
\label{fig:Fig10-ThermalConductivity1}
\end{figure*}

\subsection{Thermal conductivity}\index{SmB$_6$!thermal conductivity}

Thermal conductivity ($\kappa$) measurements are a valuable way of probing possible fermionic charge-neutral excitations, which carry entropy and would contribute to a finite residual term, $\kappa_0/T$, in the $T=0$ limit. Previous experiments performed almost 40 years ago concluded that the thermal conductivity of SmB$_6$ above 1.5~K is dominated by phonons, i.e., $\kappa/T \propto T^{2}$ \cite{Fis_Flachbart1982}, but this scenario has been revisited recently in experiments performed at much lower temperatures and much higher magnetic fields.

The first contemporary report was performed on flux-grown SmB$_6$ single crystals in a dilution refrigerator with fields to 14.5~T \cite{Fis_Xu2016}. The thermal conductivity was measured on a (100) surface using a standard four-wire steady-state method with two RuO$_2$ chip thermometers. Figure \ref{fig:Fig10-ThermalConductivity1}a displays thermal conductivity data down to 0.1~K in zero field and at 14.5~T. By fitting the zero-field data to $\kappa/T=a+bT^{\alpha-1}$, one extracts a residual term $\kappa_0/T=-0.003\pm 0.004$~mW\,K$^{-2}$cm$^{-1}$, which is zero within the experimental error bar. The contribution from surface states\index{surface states!in SmB$_6$}\index{SmB$_6$!surface states} is estimated to be two orders of magnitude smaller than this error bar and therefore negligible.

As shown in Fig.~\ref{fig:Fig10-ThermalConductivity1}\,(a) (bottom panel), $\alpha = 2.94$, in agreement with the expected phonon contribution. Finally, applied magnetic fields have very little effect on the thermal conductivity, and the residual term $\kappa_0/T$ remains negligible. Xu \textit{et al.} therefore conclude that thermal conductivity measurements do not support fermionic charge-neutral quasiparticles. One possible scenario raised by Xu \textit{et al.} to explain the bulk quantum oscillation results discussed above is spatial inhomogeneity below a percolation threshold.

An independent report on floating-zone crystals, however, reveals a finite zero-field $\kappa_0/T$, which increases as a function of applied field as shown in Fig.~\ref{fig:Fig10-ThermalConductivity1}\,(b) \cite{Fis_Hartstein2018}. Hartstein \textit{et al.} argue that phonons are unlikely to be the origin of this behavior because the phonon thermal conductivity is at its maximum in the boundary scattering limit. The similarity between the field dependent thermal conductivity of SmB$_6$ and that of organic insulator $\kappa$-(BEDT-TTF)$_2$Cu$_2$(CN)$_3$ is argued to be evidence for a neutral Fermi surface hosting excitations that transport heat but not charge.

\begin{figure*}[!t]
\centerline{\includegraphics[width=1.025\columnwidth]{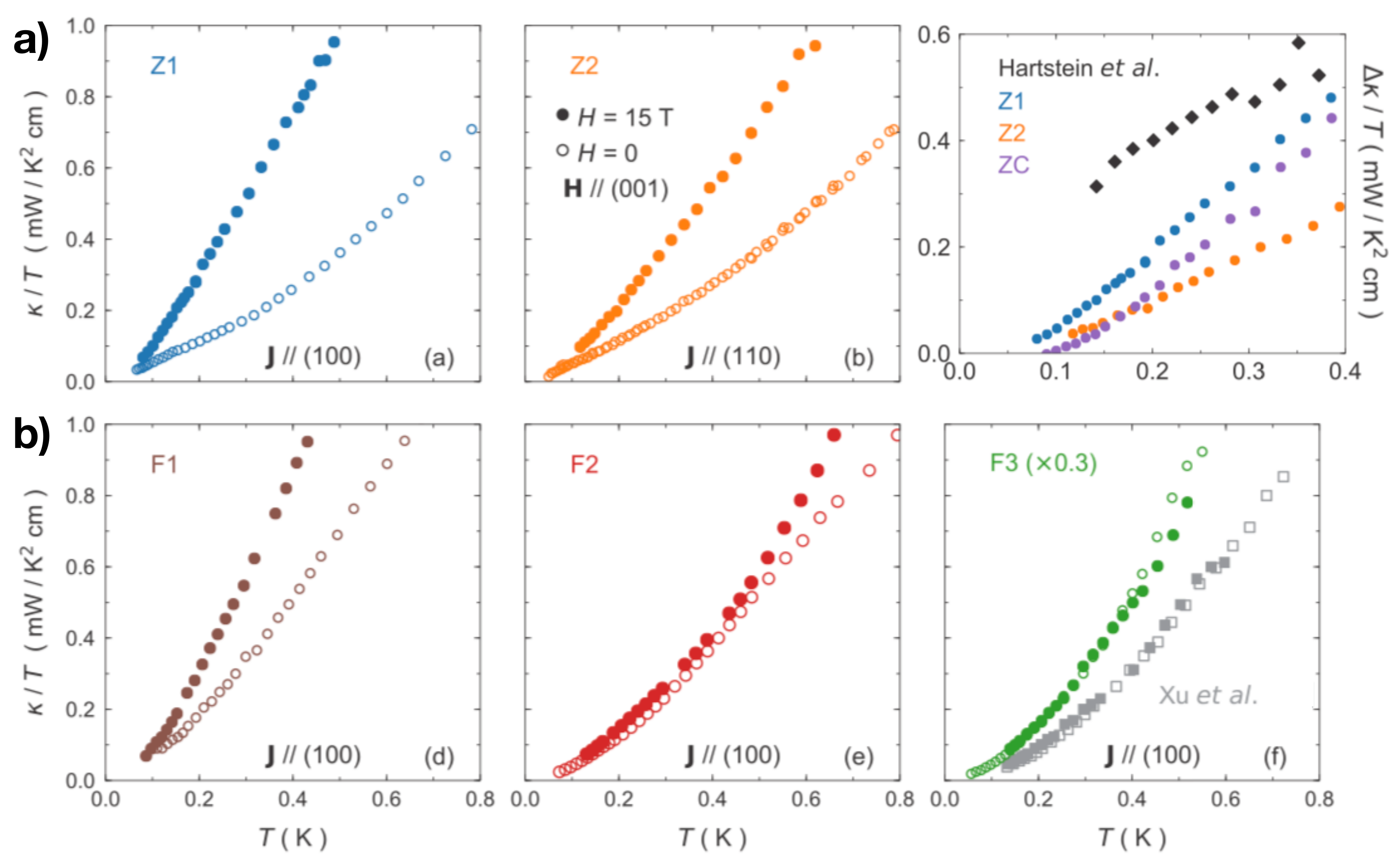}}\vspace{-2pt}
\caption{Temperature dependence of the thermal conductivity for (a)~floating-zone and (b)~flux-grown SmB$_6$ crystals. Reproduced from Boulanger \textit{et al.} \cite{Fis_Boulanger2018}. The references to Hartstein \textit{et al.} in (a) and Xu \textit{et al.} in (b) correspond to Refs.~\cite{Fis_Hartstein2018} and \cite{Fis_Xu2016}, respectively.}
\label{fig:Fig11-ThermalConductivity2}
\end{figure*}

A recent report by a third group sheds light on this controversy by measuring field-dependent thermal conductivity down to 70~mK on a variety of single crystals grown by both flux-grown and floating-zone techniques \cite{Fis_Boulanger2018}. The authors confirm the absence of a residual term $\kappa_0/T$ on six different samples in both zero field and at high magnetic fields, as summarized in Fig.~\ref{fig:Fig11-ThermalConductivity2}. A large field-induced enhancement of $\kappa$, however, is observed for all floating zone samples as well as two out of three flux-grown samples. Importantly, this sample dependence points to an extrinsic origin.

In addition, the field-induced enhancement of $\kappa$ is systematically smaller in flux-grown samples, and the behavior of sample $F3$ is in agreement with the first report discussed above \cite{Fis_Xu2016}. Generally speaking, the field dependence of the thermal conductivity is consistent with two distinct mechanisms: low-energy magnetic excitations (e.g. magnons or spinons) or phonons scattered by a field-dependent contribution (e.g. spin fluctuations or magnetic impurities).\index{impurities} The authors conclude that their data is in line with the latter scenario as magnetic impurities may significantly affect the phonon thermal conductivity in insulators even at a 1\% level. The application of magnetic field splits the energy levels of the magnetic impurities,\index{impurities} which causes a mismatch between the phonon energy and the impurity level splitting. To test their interpretation, the authors reduced the cross-sectional area of one of the floating zone samples, which makes the boundary-limited phonon conductivity smaller. According to their expectation, $\kappa/T^{2}$ decreases with decreasing cross section because the magnetic field gaps out the zero-field scattering mechanism, and the thermal conductivity is then set by the boundary limit. For more details on the analysis, including the high-temperature regime, please refer to \cite{Fis_Boulanger2018}. It is also worth noting that, though there is no evidence for fermionic neutral quasiparticles, they may be thermally decoupled from the measurable, heat-carrying phonons.\index{charge-neutral quasiparticles|)}\index{Fermi surface!charge neutral|)}

\subsection{Specific heat}\index{SmB$_6$!specific heat}
\label{Fisk-bulk-cp}

\begin{figure*}[!t]
\includegraphics[width=\columnwidth]{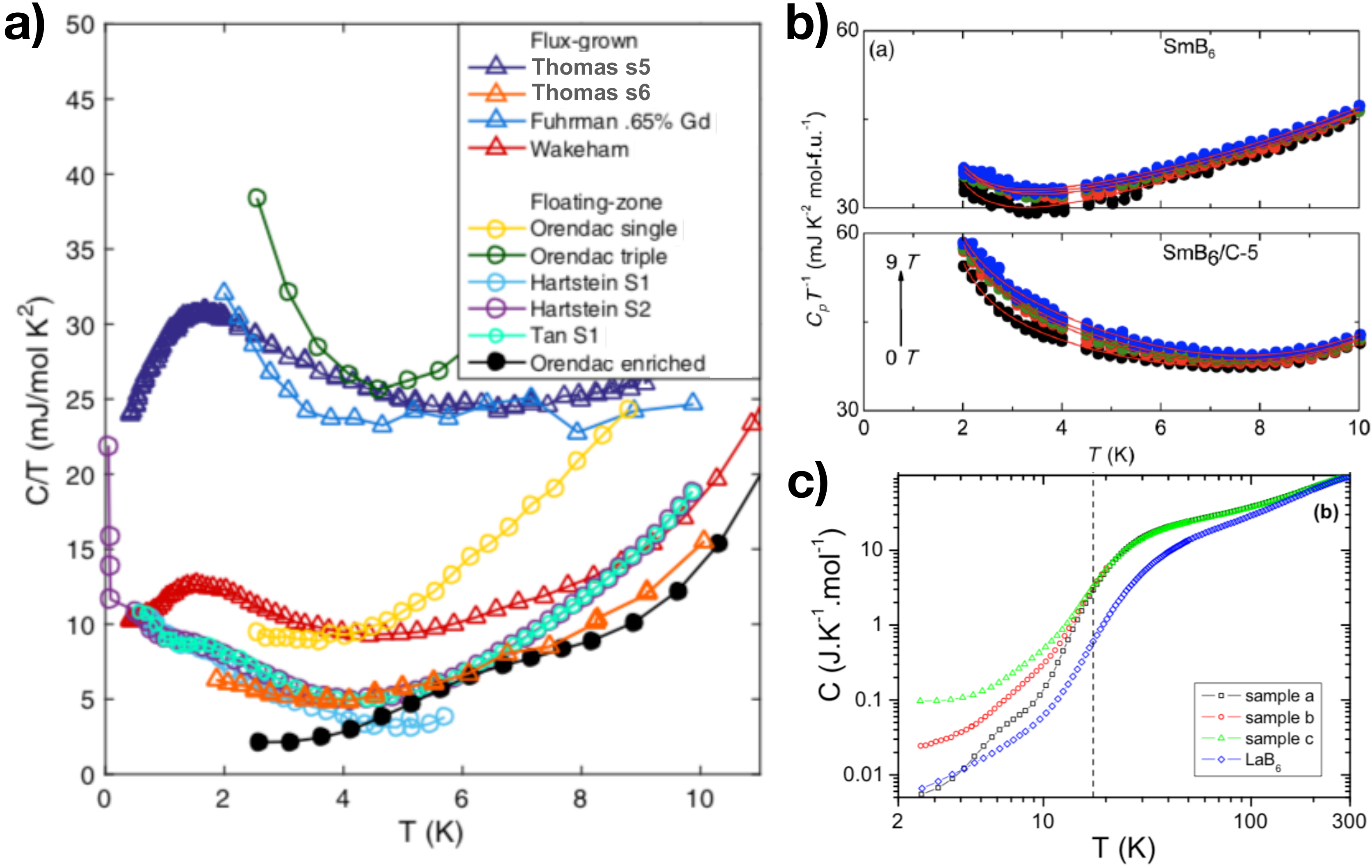}
\caption{(a)~A compilation of zero-field specific heat data as a function of temperature for SmB$_6$ crystals grown by different techniques \cite{Fis_Thomas2019, Fis_Fuhrman2018a, Fis_Wakeham2016a, Fis_Orendac2017, Fis_Tan2015, Fis_Hartstein2018}. Reproduced from Thomas \textit{et al.} \cite{Fis_Thomas2019}. (b)~Zero-field specific heat data of carbon-doped floating-zone SmB$_6$. Reproduced from Phelan \textit{et al.} \cite{Fis_Phelan2014}. (c)~Zero-field specific heat data of floating-zone LaB$_6$ and SmB$_6$. Sample a was grown with isotopically enriched elements, sample b with natural impurities and sample c with natural impurities\index{impurities} and triply melted. Reproduced from Orendac \textit{et al.} \cite{Fis_Orendac2017}.}
\label{fig:Fig12-specific-heat}
\end{figure*}

Figure~\ref{fig:Fig12-specific-heat}\,(a) shows a compilation of specific heat data on a variety of samples grown using different methods \cite{Fis_Thomas2019, Fis_Fuhrman2018a, Fis_Wakeham2016a, Fis_Orendac2017, Fis_Tan2015, Fis_Hartstein2018}. Similar to the ac and thermal conductivity measurements discussed above, specific heat measurements are sensitive to impurities\index{impurities} that may not percolate in dc resistivity measurements. In fact, the large variation of the residual specific heat in Fig.~\ref{fig:Fig12-specific-heat}\,(a) points to an extrinsic origin. The lowest specific heat magnitude reported to date was obtained on a crystal grown with isotopically enriched $^{154}$Sm and $^{11}$B \cite{Fis_Orendac2017}  providing further evidence that the broad feature centered near 1.5~K is caused by naturally occurring impurities.\index{impurities} A residual feature, however, remains at about 7~K, and we will come back to this energy scale in section \ref{Fisk-bulk-nmr}.

Rare-earth purification is challenging and expensive. Rare-earth elements obtained from Ames Laboratory are one of the purest available sources, but Ames samarium is still 99.99\% pure at best. Typically, europium, erbium, lanthanum and iron can be found at several ppm level. In agreement with this reasoning, enhanced low-temperature specific heat values are observed in flux-grown samples intentionally doped with gadolinium. In addition, specific heat data from samples with distinct Gd concentrations can be scaled to follow the same power law at low temperatures \cite{Fis_Fuhrman2018a}. This scaling was argued to be consistent with the Kondo impurity model. We note that enhanced values of the low-temperature specific heat are also observed in floating-zone samples doped with carbon \cite{Fis_Phelan2014}. Further, floating-zone samples that are doubly or triply melted display similar specific heat enhancement, though in this case it is less clear what exact type of defects one would encounter \cite{Fis_Orendac2017}.

Attempts to fit the specific heat above 2~K have been made by independent groups typically including a combination of the standard electronic ($\gamma T$) and lattice ($\beta T^{3}$) contributions as well as an exchange-enhanced paramagnetic term ($\alpha T^{3}\mathrm{ln}(T/T^{*})$) and the Schottky anomaly ($AT^{-2}$) \cite{Fis_Phelan2014, Fis_Orendac2017}. Phelan~\textit{et al.} showed that the phonon and exchange terms follow a universal curve with a characteristic temperature $T^{*}=17$~K associated with Kondo hybridization. Below 2~K, however, additional contributions come into play. Previous specific heat measurements performed in dilution refrigeration temperatures  successfully explained the \textit{difference} between in-field and zero-field measurements by calculating the contributions from nuclear magnetic moments of $^{147}$Sm, $^{149}$Sm, $^{10}$B and $^{11}$B isotopes as well as 300~ppm of magnetic impurities~\cite{Fis_Flachbart2006}.\index{impurities} A quantitative understanding of the anomalous bulk behavior of SmB$_6$ even at zero field, however, remains unavailable \cite{Fis_Stankiewicz2019, Fis_Wakeham2016a}.

\subsection{Raman spectroscopy}\index{SmB$_6$!Raman spectra|(}\index{Raman scattering!by phonons|(}
\label{Fisk-bulk-raman}

\begin{figure*}[!t]
\centerline{\includegraphics[width=0.9\columnwidth]{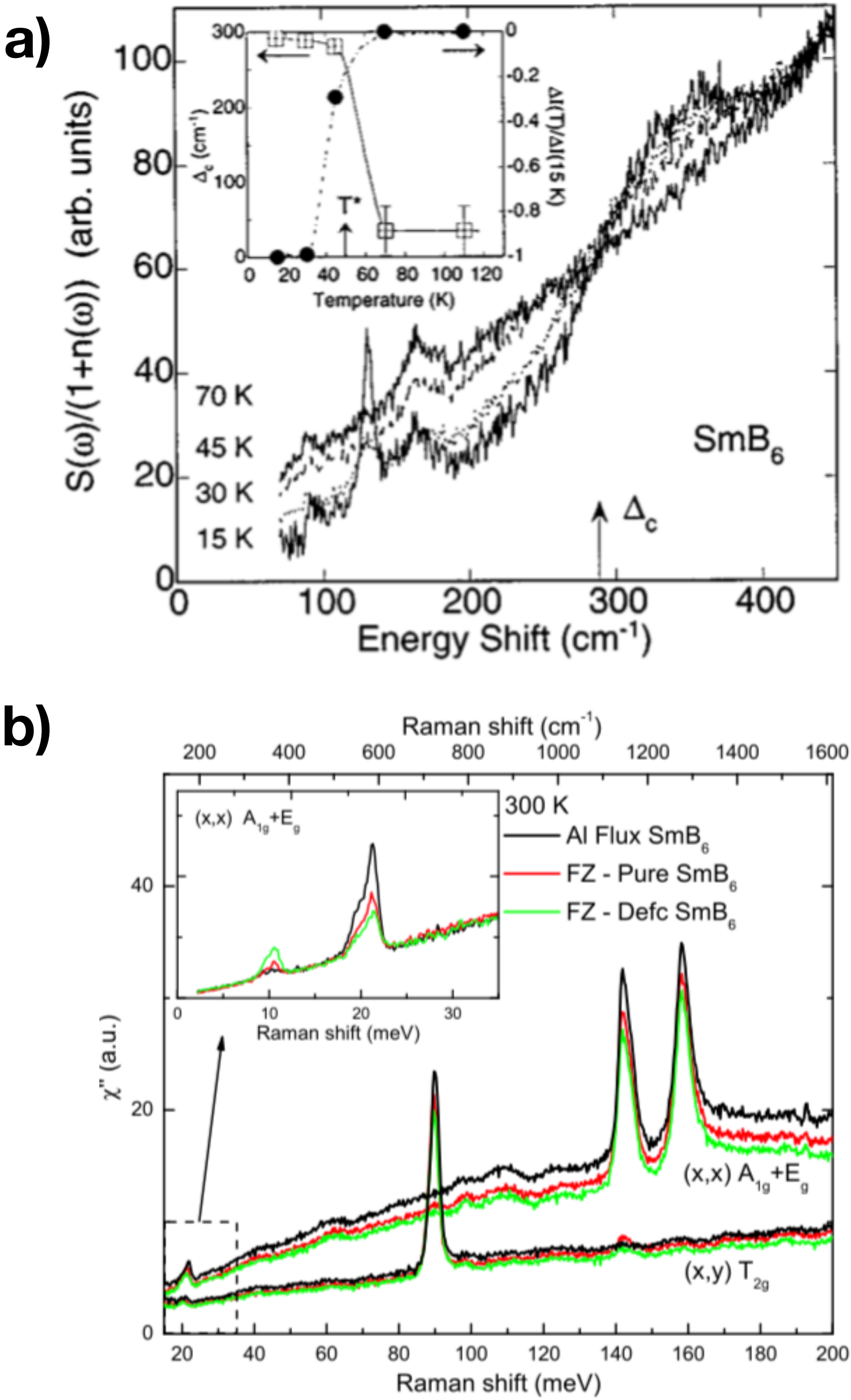}}
\caption{Raman scattering response function of a flux-grown SmB$_6$ single crystal at various temperatures (panel a). Reproduced from Nyhus~\textit{et al.}~\cite{Fis_Nyhus1995}. Raman scattering response function of of both flux-grown and floating-zone crystals at room temperature (panel b). Reproduced from Valentine \textit{et al.} \cite{Fis_Valentine2016}.\vspace{-2em}}
\label{fig:Fig13-raman}
\end{figure*}

Now we turn our attention to spectroscopic techniques. Raman spectroscopy is based on the inelastic scattering of light in a material, providing a direct means of probing the development of an energy gap. This technique also provides energy and symmetry information about other low-frequency excitations including crystal-field excitations, excitons, and anomalous phonons. Evidence for an energy gap in SmB$_6$ was observed in early reports of the temperature dependence of the Raman continuum scattering, as shown in Fig.~\ref{fig:Fig13-raman}\,(a) \cite{Fis_Nyhus1995}. The electronic scattering intensity shifts abruptly below \mbox{$T^{*}\approx50$\,--\,60~K}, which corresponds to the temperature at which the magnetic susceptibility peaks. The shift in spectral weight is centered about an isosbestic point at 290~cm$^{-1}$, an energy substantially larger than the energy scale over which the gap appears in the scattering response. A number of theoretical models were proposed to account for electrical transport as well as Raman data \cite{Fis_Nyhus1997}; however, the discovery of non-Ohmic behavior arising from a conducting surface state in SmB$_6$ discussed above makes analysis not taking this into account of limited value.

In the past few years, the Raman response of SmB$_6$ was revisited, and a timely sample dependence investigation was performed \cite{Fis_Valentine2016}. Figure~\ref{fig:Fig13-raman}\,(b) shows the room temperature Raman spectra of three distinct single crystals: one crystal grown by the flux technique (Al flux SmB$_6$), a second crystal grown by the floating zone technique close to being stoichiometric (FZ SmB$_6$ - Pure), and a third crystals also grown by the floating zone technique but less stoichiometric (FZ SmB$_6$ -- Defc). As we mentioned above, an increase in the number of Sm vacancies is observed along the length of a floating zone crystal rod due to the vaporization and has been characterized by a systematic decrease in lattice parameters.

Three symmetry-allowed phonons are observed in Fig.~\ref{fig:Fig13-raman}\,(b), in agreement with previous reports \cite{Fis_Morke1981}. These narrow peaks are related to distortions of the B$_6$ octahedra as follows: $T_{2g}$ phonon at 89.6~mev (723~cm$^{-1}$), $E_g$ phonon at 141.7~meV (1143~cm$^{-1}$), and $A_{1g}$ phonon at 158.2~meV (1277~cm$^{-1}$). At much lower energies, however, two additional features appear at 10~meV and 21~meV, which are not allowed by the cubic space group in first-order Raman scattering. The latter feature has been previously assigned to a two-phonon scattering mode \cite{Fis_Nyhus1995}. The former feature, however, has been recently attributed to a finite-momentum scattering from acoustic phonons due to local symmetry breaking induced by the presence of Sm defects \cite{Fis_Valentine2016}. Neutron scattering experiments show a relatively flat dispersion of the acoustic phonon branches, in agreement with the narrow line width at 10~meV \cite{Fis_Alekseev1989}. Further evidence for this scenario comes from the 50\% increase in spectral weight at 10~meV in the most deficient floating-zone sample, which is estimated to have only about 1\% of Sm vacancies. A significant impurity-driven enhancement is also observed in the specific heat of these samples \cite{Fis_Valentine2018}.
\index{SmB$_6$!Raman spectra|)}\index{Raman scattering!by phonons|)}

\subsection{X-ray, neutron and M\"{o}ssbauer spectroscopies}\index{SmB$_6$!M\"{o}ssbauer spectroscopy|(}
\label{Fisk-bulk-neutron}

\begin{figure*}[!t]
\includegraphics[width=\columnwidth]{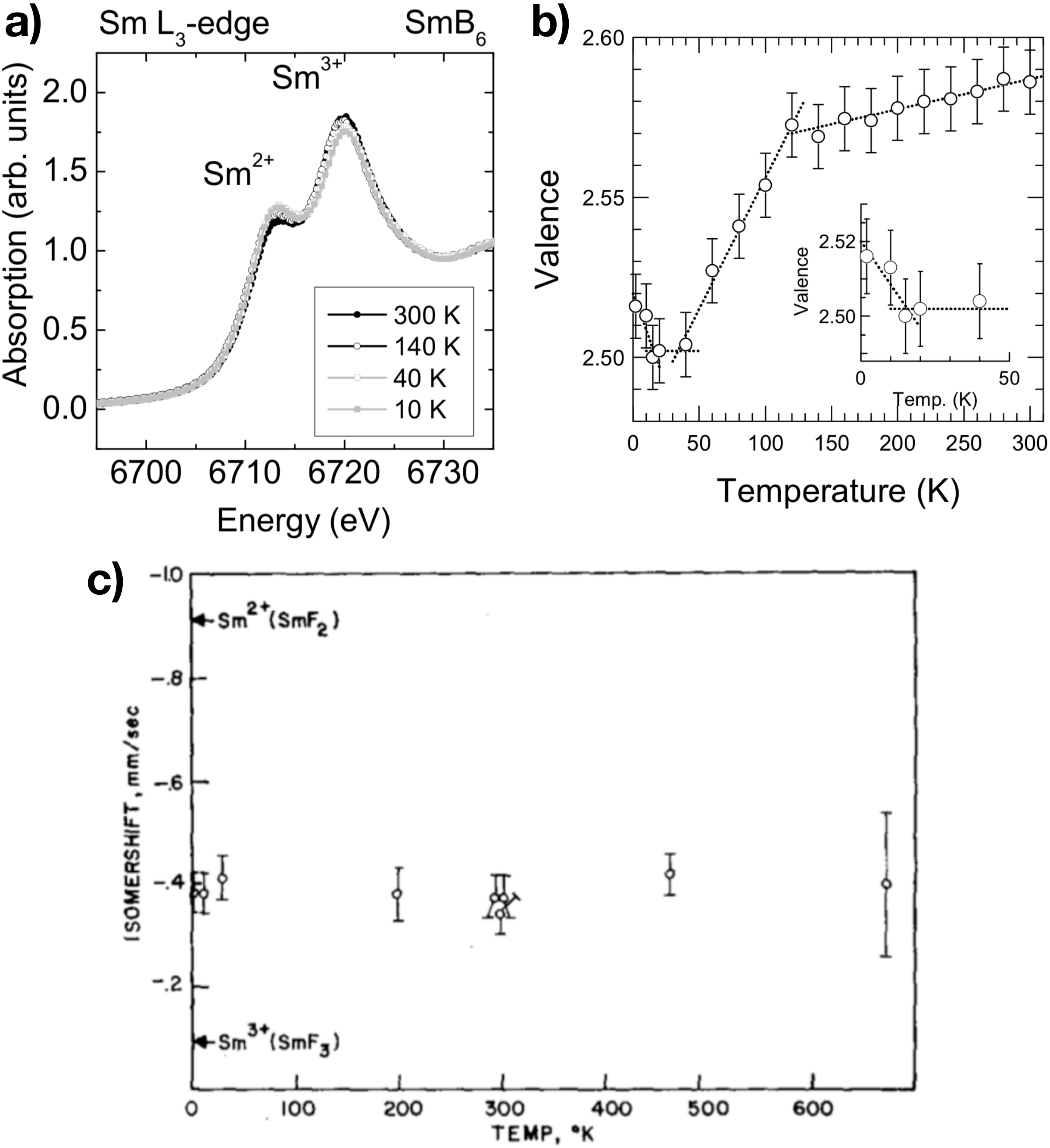}\vspace{-2pt}
\caption{(a)~Sm $L_3$-edge x-ray absorption spectra at different temperature. Reproduced from Mizumaki \textit{et al.} \cite{Fis_Mizumaki2009}.
(b)~Average Sm valence as a function of temperature. Reproduced from Mizumaki~\textit{et al.}~\cite{Fis_Mizumaki2009}.
(c)~Isomer shift as a function of temperature. Reproduced from Cohen \textit{et al.} \cite{Fis_Cohen1970}.}
\label{fig:Fig14-mossbauer}
\end{figure*}

Early $L_3$-edge x-ray absorption experiments\cite{Fis_Vainshtein1965} as well as $^{149}$Sm M\"{o}ssbauer resonance measurements \cite{Fis_Cohen1970, Fis_Cohen1970b} estimated the valence\index{SmB$_6$!mixed valence} of Sm to be +2.6 at room temperature.\index{SmB$_6$!x-ray absorption spectroscopy} X-ray absorption spectra display two separate peaks corresponding to Sm$^{2+}$ and Sm$^{3+}$ \cite{Fis_Mizumaki2009}. The spectra evolve with temperature as shown in Fig.~\ref{fig:Fig14-mossbauer}\,(a,\,b), revealing that the valence of Sm decreases with decreasing temperature. The temperature dependence of the lattice parameters \cite{Fis_Tarascon1980} combined with negative thermal expansion and anomalous elastic constants at low temperatures \cite{Fis_Mandrus1994, Fis_Nakamura1991} also indicate valence changes as a function of temperature. M\"{o}ssbauer resonance, however, only contains a single-line spectrum, and no change in the Sm valence as a function of temperature is detected by isomer shift measurements, Fig.~\ref{fig:Fig14-mossbauer}\,(c). This apparent contradiction can be resolved by taking into account the different time scales of these two measurements. X-ray absorption is a faster measurement (10$^{-12}$~s) than M\"{o}ssbauer resonance (10$^{-8}$~s), indicating that the valence of Sm fluctuations at about 10$^{-8}$~s. An alternative explanation, however, has been recently proposed based on the boron-dimer model discussed in section \ref{Fisk-theory} \cite{Fis_Robinson2019}. The authors predict a constant ratio of $f$-shell occupations, which is in turn probed by M\"{o}ssbauer spectroscopy, and a variable $d$-orbital occupation, which cannot be probed by M\"{o}ssbauer measurements. We note that the Sm valence obtained from the above Sm $L_3$ x-ray absorption spectroscopy results has been reproduced by several independent groups as well as by distinct methods including x-ray photoelectron spectroscopy  \cite{Fis_Allen1980, Fis_Hayashi2013, Fis_Lutz2016, Fis_Utsumi2017}. The M\"{o}ssbauer isomer shift result has been also recently revisited and reproduced by Tsutsui \textit{et al.} \cite{Fis_Tsutsui2016}.

Resonant x-ray emission spectroscopy as well as x-ray absorption spectroscopy\index{SmB$_6$!x-ray absorption spectroscopy} also have been employed to investigate the valence\index{SmB$_6$!mixed valence} of SmB$_6$ under applied pressure.\index{SmB$_6$!under pressure} Electrical resistivity measurements under hydrostatic conditions show that the insulating gap of SmB$_6$ vanishes at $P_{\rm c} = 10$~GPa accompanied by magnetic order \cite{Fis_Derr2008}. This critical pressure is dependent on the hydrostaticity of the pressure media and may be as low as $\sim 4$~GPa in quasi-hydrostatic environments \cite{Fis_Cooley1994, Fis_Barla2005, Fis_ZhouRosa20}. X-ray spectroscopy measurements report the increase in Sm valence under pressure towards the trivalent state as the systems goes towards the antiferromagnetic\index{SmB$_6$!magnetism|(} metallic ground state \cite{Fis_Beaurepaire1990, Fis_Butch2016, Fis_Zhou2017, Fis_Chen2018, Fis_Emi2018a}. A discrepancy, however, is observed in the quantitative analysis by different groups. For example, Zhou \textit{et al.} report a valence very close to $3+$ at 20~GPa \cite{Fis_Zhou2017}, whereas other groups argue for a finite divalent character to 35~GPa \cite{Fis_Butch2016, Fis_Chen2018, Fis_Emi2018}.

Sm $M$-edge x-ray absorption and x-ray magnetic circular dichroism\index{SmB$_6$!x-ray magnetic circular dichroism} measurements have been simultaneously performed on the surface and in the bulk of SmB$_6$, respectively. Phelan \textit{et al.} confirm the presence of Sm$^{2+}$ and Sm$^{3+}$ in the bulk of floating-zone crystals; however, the polished surface is shown to contain mostly Sm$^{3+}$ with a net magnetic moment of 0.09~$\mu_{\rm B}$ at $T=10$~K. The authors argue that the discrepancy between the valence at the surface and in the bulk could generate band bending at the interface. Though Chen \textit{et al.} also report a higher Sm valence at the surface of powdered floating-zone SmB$_6$, the extracted value of $\nu = 2.7$ is below the trivalent state \cite{Fis_Chen2018}. Employing the same techniques on vacuum-cleaved samples reveal that both divalent and trivalent Sm are present on the surface of SmB$_6$, indicating that polishing has a significant effect on the surface valence \cite{Fis_Fuhrman2019}. In addition, Fuhrman \textit{et al.} report that the Sm$^{3+}$ magnetic dipole moment antialigns with an applied magnetic field below $T=75$~K and suggest that Sm$^{3+}$ couples antiferromagnetically to large-moment paramagnetic impurities\index{impurities} known to be present in the samples. Sm $N$-edge x-ray absorption spectroscopy has also been performed on flux-grown SmB$_6$ \cite{Fis_He2017}. He \textit{et al.} show that the Sm$^{3+}$ contribution on a cleaved surface increases irreversibly with time as it ages in an ultra-high-vacuum chamber. In fact, soft x-ray absorption and reflectometry measurements on floating-zone crystals reveal that a cleaved $(001)$ surface of SmB$_6$ undergoes significant valence and chemical reconstruction\index{surface reconstruction} as a function of time. The final surface is argued to be boron-terminated with a Sm$^{3+}$ subsurface region. At room temperature, this reconstruction takes less than two hours, whereas below 50~K it takes about a day \cite{Fis_Zabolotnyy2018}. These results shed light on discrepancies observed in scanning tunneling spectroscopy measurements discussed in section \ref{Fisk-surface-stm}.

Recent non-resonant and resonant x-ray scattering\index{SmB$_6$!x-ray scattering spectroscopy} experiments provide a powerful tool to determine the ground state and crystal field splitting of SmB$_6$, respectively. On one hand, core-level non-resonant scattering in the hard x-ray region is a bulk sensitive spectroscopic technique performed at large momentum transfer. The scattering function therefore provides information on higher multipole terms and allows the determination of the ground state wave function even in cubic compounds \cite{Fis_Sundermann2018}. On the other hand, $M_{4,5}$-edge ($3d \rightarrow 4f$) resonant inelastic x-ray scattering is both element- and configuration-sensitive and allows the determination of the crystal field splitting of the Sm$^{3+}$ Hund's rule ground state \cite{Fis_Amorese2019}. These two spectroscopic techniques find that the $J=5/2$ multiplet of Sm $f^{5}$ splits into a $\Gamma_{8}$ quartet ground state and a $\Gamma_{7}$ first excited doublet at $20$~meV. It is worth noting that this crystal field splitting agrees well with the expected splitting from the extrapolation of crystal field parameters obtained within the $R$B$_6$ series ($R$ is a rare-earth element).

Inelastic neutron scattering\index{SmB$_6$!INS spectroscopy} has been also employed to shed light on this problem. The strong neutron absorption of both samarium and boron, combined with the presence of hybridization and two Sm configurations, makes these measurements challenging. Nevertheless, early inelastic neutron scattering experiments using double isotopic samples, $^{154}$Sm$^{11}$B, identify broad intermultiplet transitions at about $36$~meV and $130$~meV as well as a narrow low-energy excitation at the $R$ point [($\frac{1}{2}\frac{1}{2}\frac{1}{2}$)] centered at $14$~meV only observed at temperatures below $100$~K \cite{Fis_Alekseev1989, Fis_Alekseev1992, Fis_Alekseev1993, Fis_Alekseev1995, Fis_Alekseev2010}. This low-temperature feature does not follow the localized $f$-electron form factor and was therefore argued to be caused by the mixed-valence state of Sm and exciton states within the gap.

\begin{figure*}[t!]
\includegraphics[width=\columnwidth]{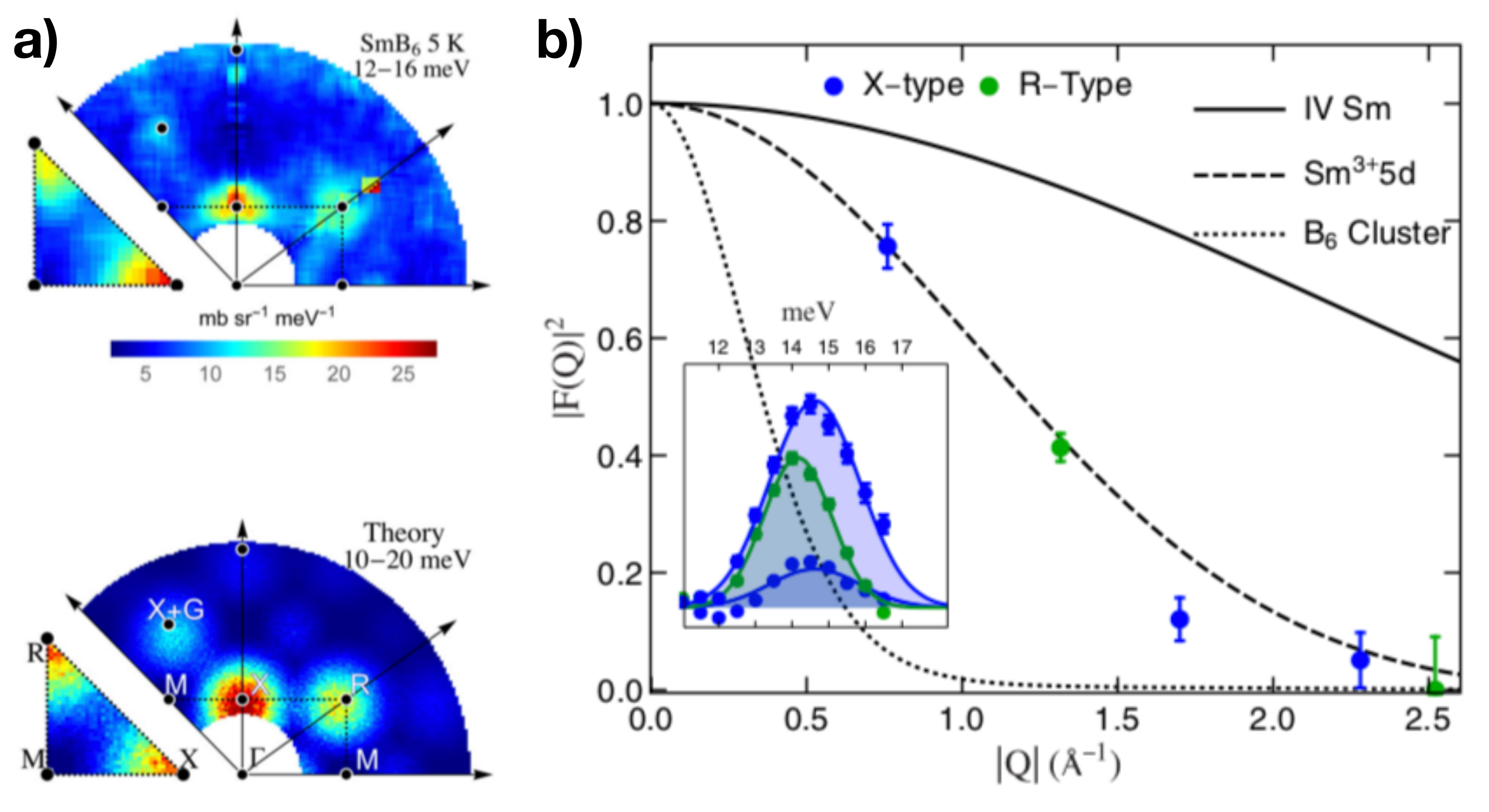}
\caption{(a)~Experimental (top) and theoretical (bottom) energy integrated neutron scattering intensity of floating-zone $^{154}$Sm$^{11}$B at high-symmetry planes. (b)~Comparison between the squared magnetic form factor of different scattering centers (lines) and the experimental integrated neutron scattering intensity (symbols) as a function of momentum $\mathbf{Q}$. Reproduced from Fuhrman \textit{et al.}~\cite{Fis_Fuhrman2015}.
\index{SmB$_6$!INS spectroscopy}}
\label{fig:Fig15-INS}
\end{figure*}

Recent inelastic neutron scattering\index{SmB$_6$!INS spectroscopy} measurements have revisited the low-energy, low-temperature features of SmB$_6$ in the entire Brillouin zone \cite{Fis_Fuhrman2015}. Fuhrman~\textit{et al.} show that the narrow ($>2$~meV) resonant mode at $14$~meV is also intense near the $X$ point ($\frac{1}{2}00$), see Fig.~\ref{fig:Fig15-INS}\,(a), and, though much weaker, goes beyond the first zone. As a result, the form factor is mapped out and shown to unexpectedly follow the $5d$ electron form factor, Fig.~\ref{fig:Fig15-INS}\,(b), indicating a critical role of such orbitals in the exciton formation and providing strong constraints to realistic theories. Fuhrman \textit{et al.} propose a minimal phenomenological model with third neighbor hopping,\index{hopping} which allows for $5d$-electron ``pseudo-nesting'' to enhance the generalized susceptibility that appears in the inelastic neutron scattering measurements through interband transitions. This phenomenological model generates a band structure with inversion pockets at $X$, in agreement with the topological Kondo insulator picture.\index{topological Kondo insulator}

The spin-exciton may also shed light on the presence of impurities in SmB$_6$ as coupling of the exciton to a fermionic density of states at $E_{\rm F}$ would cause a finite relaxation rate in inelastic neutron scattering measurements. A nearly temperature independent lifetime, however, is observed from 15 to 3.5~K, indicating no coupling to additional energy scales below the 14~meV spin-exciton mode. In addition, no indication of magnetism was observed in the low-energy spectrum, and an upper bound fluctuating moment of 0.05(2)~$\mu_{\rm B}^{2}$ was estimated. The authors conclude that magnetic impurities\index{impurities} in SmB$_6$ may be screened.
\index{SmB$_6$!M\"{o}ssbauer spectroscopy|)}

\subsection{Nuclear, electron and muon spin resonance~}
\label{Fisk-bulk-nmr}

\begin{figure*}[!t]
\includegraphics[width=\columnwidth]{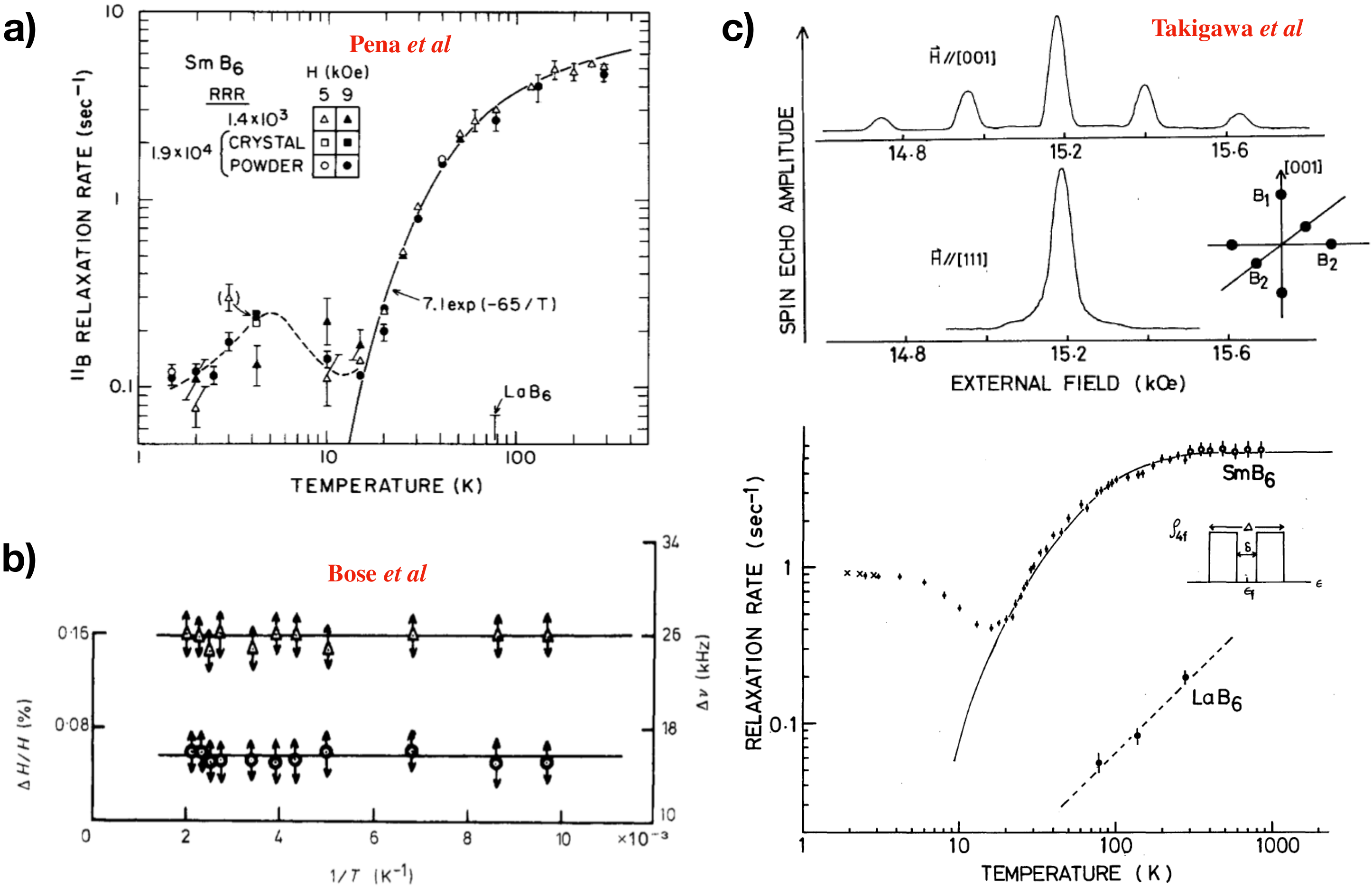}
\caption{(a)~$^{11}$B relaxation rate as a function of temperature in flux-grown powdered SmB$_6$. Reproduced from Pe\~{n}a \textit{et al.} \cite{Fis_Pena1981}. (b)~NMR linewidth as a function of inverse temperature. Reproduced from Bose \textit{et al.} \cite{Fis_Bose1980}. (c)~Spin-echo spectra at 4.2~K and 20.7~MHz (top) and $^{11}$B relaxation rate (bottom) of a SmB$_6$ single crystal. Reproduced from Takigawa \textit{et al.} \cite{Fis_Takigawa1981}.}
\label{fig:Fig16-NMRa}
\end{figure*}

Nuclear magnetic resonance\index{SmB$_6$!nuclear magnetic resonance}\index{nuclear magnetic resonance!in SmB$_6$} takes advantage of the nonzero nuclear spin moment of many stable isotopes to obtain local information on the internal field as well as spin dynamics in a material. Though it is challenging to detect Sm nuclei in SmB$_6$, resonance lines from $^{11}$B are easy to detect and allow an indirect measure of the Sm mixed valence\index{SmB$_6$!mixed valence} through the hyperfine interaction. Early nuclear magnetic resonance measurements by Pe\~{n}a \textit{et al.} showed that the $^{11}$B spin lattice relaxation above 15~K is activated with an energy gap of about 6~meV, in rough agreement with the transport gap (i.e., $4$~meV) \cite{Fis_Pena1981, Fis_Pena1981a}. An anomalous peak, however, emerges at about 5~K as shown in Fig.~\ref{fig:Fig16-NMRa}\,(a). The authors state that evaluation of models for this anomaly is hindered by the lack of precise defect characterization in nearly pure SmB$_6$, and this statement remains accurate to date. Pe\~{n}a \textit{et al.} argue that the maximum may be related to a decrease in fluctuation amplitude at low temperatures, which in turn could be caused by Sm$^{3+}$ formation near Sm vacancies.

A second $^{11}$B nuclear magnetic resonance\index{nuclear magnetic resonance!in SmB$_6$} experiment reported constant Knight shift and line width data from 100~K to 480~K \cite{Fis_Bose1980}. Bose \textit{et al.} ascribed this result to a temperature independent valence fluctuation of the Sm ion owing to the fast fluctuation rate ($<10^{13}$~s$^{-1}$), Fig.~\ref{fig:Fig16-NMRa}\,(b). A third set of $^{11}$B nuclear magnetic resonance measurements were carried out by Takigawa \textit{et al.} at low fields and at temperatures between 2.5~K and 850~K, as shown in Fig.~\ref{fig:Fig16-NMRa}\,(c) \cite{Fis_Takigawa1981}. The temperature dependence of the relaxation rate above 20~K was modeled taking into account a 50~K gap, and they attribute the increase in $1/T_1$ below 15~K to the Wigner crystal scenario.\index{Wigner crystallization} These data are similar to those reported by Pe\~{n}a \textit{et al.}; however, $1/T_1$ plateaus below 5~K, instead of decreasing with decreasing temperature, which could be due to an anisotropy in the relaxation rate. It is worth noting that Pe\~{n}a \textit{et al.} powdered a collection of flux-grown single crystals for the experiments, whereas the sample preparation procedure was not described by Bose \textit{et al.} and Takigawa \textit{et al.}

\begin{figure*}[!t]
\includegraphics[width=\columnwidth]{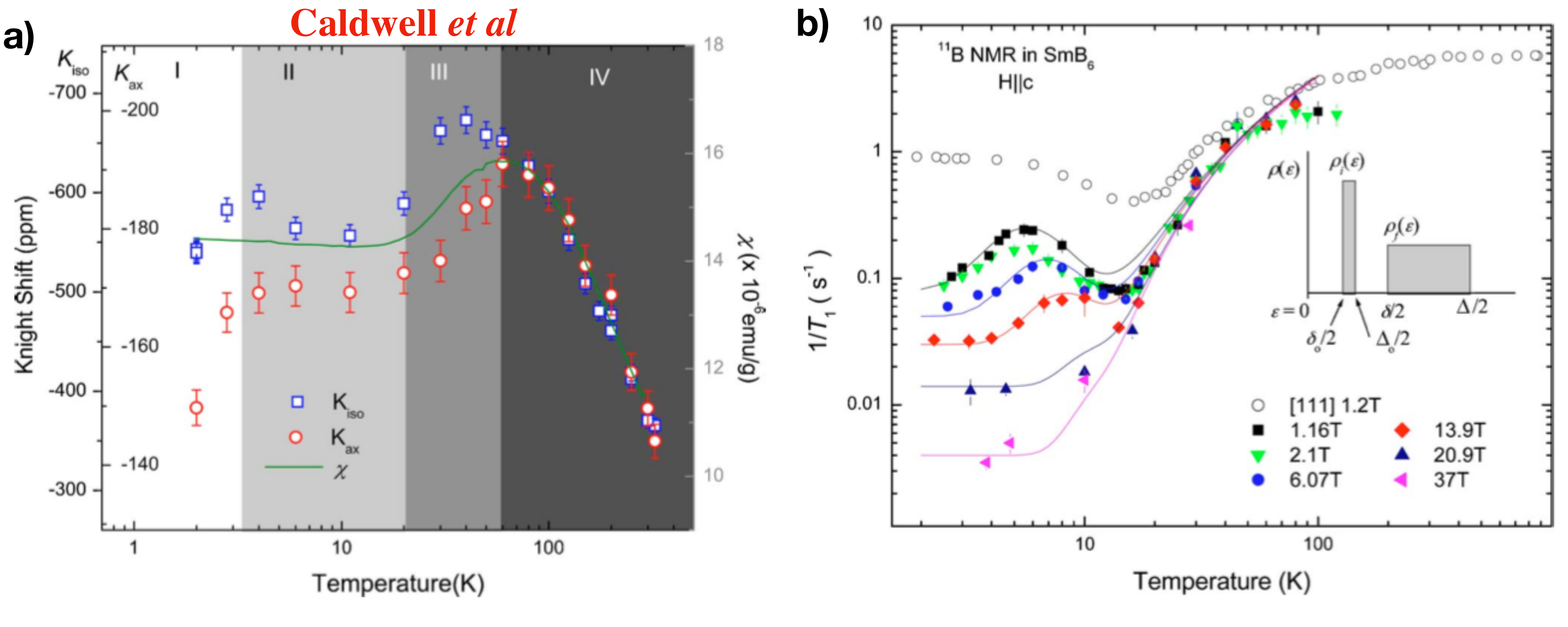}
\caption{(a)~NMR Knight shift as a function of temperature. (b)~$^{11}$B relaxation rate as a function of temperature in flux-grown SmB$_6$ single crystals for various applied fields along the $c$-axis. Reproduced from Caldwell \textit{et al.} \cite{Fis_Caldwell2007}.}
\label{fig:Fig16-NMRb}
\end{figure*}

In 2007, nuclear magnetic resonance\index{nuclear magnetic resonance!in SmB$_6$} measurements in flux-grown SmB$_6$ single crystals were reported in high magnetic fields to 37~T \cite{Fis_Caldwell2007}. Caldwell \textit{et al.} confirm the opening of a gap in the density of states below 100~K and the presence of in-gap states, which dominate the nuclear relaxation below about 10~K, as shown in Fig.~\ref{fig:Fig16-NMRb}. The application of high magnetic fields suppress this in-gap contribution, whereas the gap remains open to 37~T. The authors model their data with a gapped density of states containing a symmetric rectangular distribution of in-gap states as shown in Fig.~\ref{fig:Fig16-NMRb}\,(b). This simplified density of states is able to fit the experimental data, though the nature of the field-dependent in-gap states remains unclear. The authors argue that an impurity band caused by boron vacancies is not consistent with the significant field dependence of the data. We note that it could be useful to revisit these measurements in light of the recent information on surface states,\index{surface states!in SmB$_6$}\index{SmB$_6$!surface states} spin excitons\index{spin exciton} and magnetic impurities.\index{impurities} For a recent theoretical perspective on this issue, please refer to \cite{Fis_Schlottmann2014}.

Nuclear magnetic resonance\index{nuclear magnetic resonance!in SmB$_6$} measurements on floating-zone SmB$_6$ have been performed under hydrostatic pressures to 5~GPa \cite{Fis_Pristas2010, Fis_Pristas2011, Fis_Nishiyama2013}. The temperature dependence of the spin lattice relaxation remains activated under pressure, though the insulating gap is reduced by about 30\% at 5~GPa. As discussed in the previous section, the gap collapse is consistent with dc electrical resistivity measurements and is related to an increase of the Sm valence towards a trivalent state \cite{Fis_Beaurepaire1990, Fis_Butch2016, Fis_Zhou2017, Fis_Emi2018a}.

\begin{figure*}[!b]
\includegraphics[width=\columnwidth]{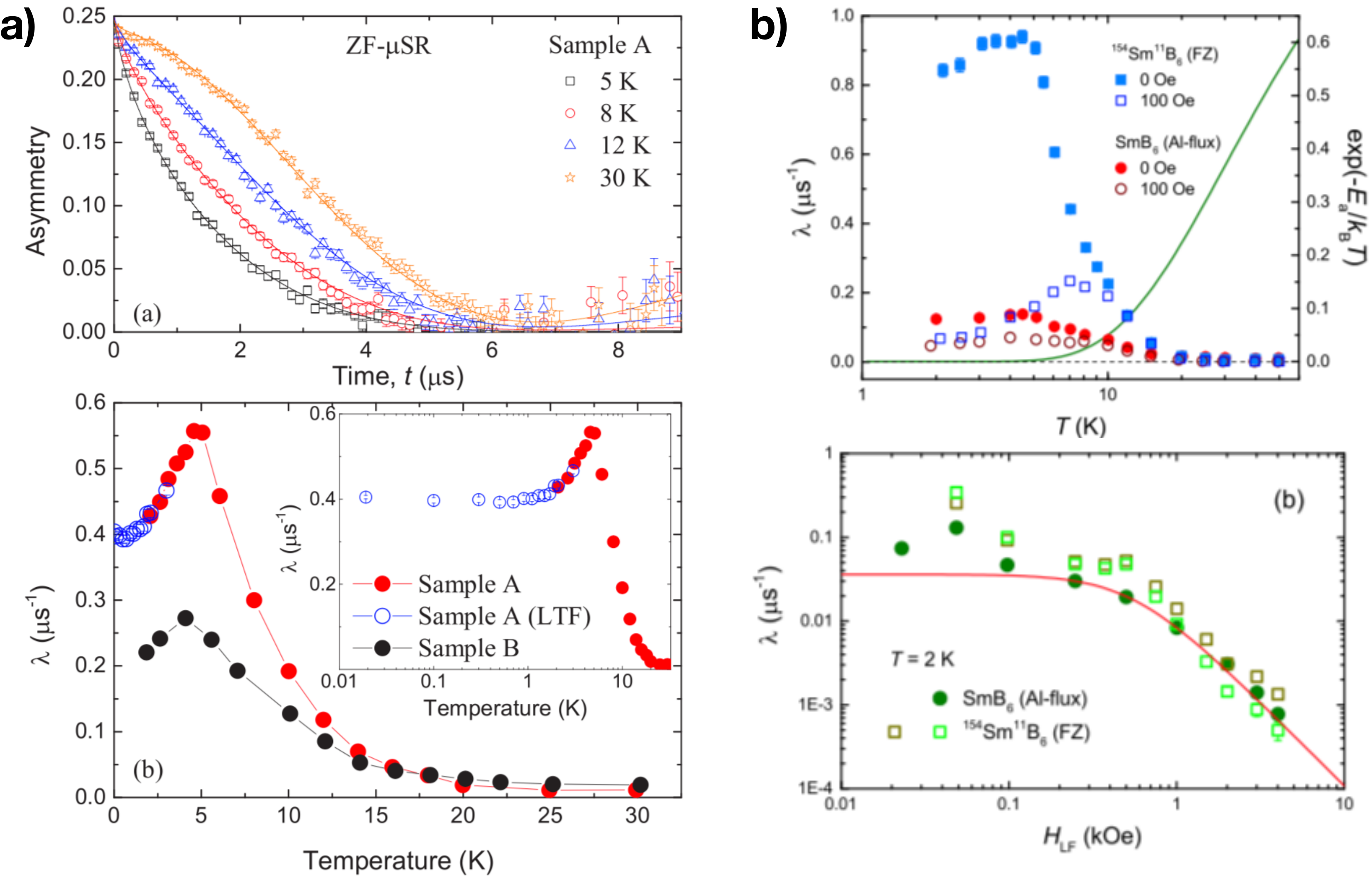}\vspace{-1pt}
\caption{(a)~Top: Zero-field muon spin resonance asymmetry signal of floating-zone SmB$_6$. Bottom: Temperature dependence of the muon spin relaxation rate of floating-zone SmB$_6$ (sample A) and flux-grown SmB$_6$ (sample B). Reproduced from Biswas~\textit{et al.}~\cite{Fis_Biswas2014}. (b)~Top: Muon spin relaxation rate at zero field (solid symbols) and 100~Oe (open symbols) of floating-zone SmB$_6$ (FZ) and flux-grown SmB$_6$ (Al flux). Bottom:~Field dependence of the relaxation rate obtained from fits of the asymmetry spectra at 2~K. Reproduced from Gheidi \textit{et al.} \cite{Fis_Gheidi2019}.\vspace{-2pt}}
\label{fig:Fig17-mSR}
\end{figure*}

As with nuclear magnetic resonance,\index{nuclear magnetic resonance!in SmB$_6$} muon spin resonance\index{SmB$_6$!muon spin relaxation}\index{muon spin relaxation} is also a sensitive local magnetic probe, which could shed light on the nature of the in-gap states in SmB$_6$. Polarized muon particles\,---\,subatomic particles similar to the electron but two hundred times heavier\,---\,are implanted in the bulk of the samples, and the asymmetry of the muon decay in time is measured. To our knowledge, the first muon spin resonance measurements in SmB$_6$ were reported in 2014 \cite{Fis_Biswas2014}. Biswas \textit{et al.} observe homogeneous magnetic field fluctuations in the bulk of SmB$_6$ below about 15~K in both floating-zone and flux-grown samples, as shown in Fig.~\ref{fig:Fig17-mSR}\,(a). The similarity between this temperature scale and those of nuclear magnetic resonance discussed above indicate a common origin. We note that no magnetic order was observed down to 19~mK. In a follow-up report, Biswas~\textit{et al.} attribute the observed magnetic excitations to spin excitons\index{spin exciton} in the bulk of SmB$_6$. These spin excitons produce local magnetic fields of about 1.8~mT fluctuating on a time scale of about 60~ns. The authors find a suppression of these fluctuations at the surface and, as a result, a small enhancement in static magnetic fields \cite{Fis_Biswas2017}. According to the authors, the difference in magnitude of the asymmetry decay between floating-zone and flux-grown samples could be coming from a nonrelaxing background signal in the small flux-grown samples or slightly different microscopic properties.

A recent report by an independent group, however, argues for the presence of sample-dependent quasi-static magnetism of extrinsic origin below about 10~K \cite{Fis_Gheidi2019}. Gheidi \textit{et al.}, however, also suggest the presence of intrinsic magnetism\index{SmB$_6$!magnetism|)} originated from two different sources, the first being an underlying low-energy weak fluctuating moment ($10^{-2}$~$\mu_{\rm B}$), and the second being consistent with a $2.6$~meV bulk magnetic excitation gap at zero field \cite{Fis_Akintola2017, Fis_Akintola2018}. The question of why this magnetic excitation is not detected by inelastic neutron scattering remains open.

Electron spin resonance\index{SmB$_6$!electron spin resonance} is also a local probe of paramagnetic ions in a materials. Though conduction-electron spin resonance has been reported in alkaline-earth hexaborides \cite{Fis_Rupp1969}, this conduction-electron signal appears to be absent in SmB$_6$. Intentional doping with paramagnetic probes is therefore required for the observation of a signal. Early electron spin resonance measurements of Er$^{3+}$ in SmB$_6$ at 4.2~K were not consistent with a purely cubic symmetry \cite{Fis_Sturm1985}. This inconsistency was ascribed to dynamical Jahn-Teller effect on the cubic $\Gamma_{8}$ crystal-field ground state \cite{Fis_Weber1985}. Recent experiments revisited this scenario and proposed that the small Er$^{3+}$ ions experience anharmonic rattling vibrations instead \cite{Fis_Lesseux2017}. Lesseux~\textit{et al.} also performed electron spin resonance measurements on SmB$_6$ crystals doped with Dy$^{3+}$ and Nd$^{3+}$, but no appreciable signal was observed.

Gd$^{3+}$ is a powerful paramagnetic probe in electron spin resonance measurements owing to its $L=0$ ground state. Early electron spin resonance measurements of Gd$^{3+}$ in SmB$_6$ revealed an unusual electron spin resonance spectrum below 5~K, which could not be explained by the $4f^{7}$ configuration \cite{Fis_Wiese1990}. Wiese~\textit{et al.} argue for a divalent $4f^{7}5d^{1}$ configuration for Gd owing to a trapped conduction electron. These experiments have been also revisited recently by Souza~\textit{et al.}, and the origin of the unusual Gd$^{2+}$-like signal is found to be an oxide impurity phase on the surface of the crystal \cite{Fis_Souza2019}. In the very dilute limit, the low-temperature fine structure of the Gd$^{3+}$ electron spin resonance spectra follows the expected cubic crystal-field environment. At high temperatures, a single resonance is observed and follows an activation behavior with a gap of about 50~K. Though these experiments have thus far probed the bulk of SmB$_6$, electron spin resonance may also be sensitive to the surface of SmB$_6$, which invites further experimental work.

\vspace{-3pt}\section{Concluding remarks}\vspace{-1pt}

The current fundamental question regarding SmB$_6$ is whether it is actually a strongly correlated topological insulator, a question which still remains to be definitively resolved. It is generally agreed that a conducting surface state exists,\index{surface states!in SmB$_6$}\index{SmB$_6$!surface states} and it appears to be sensitive to magnetic probes such as Gd. The apparent gap protection seen in the measurements carefully separating bulk from surface resistivity remains a mystery. Related to this is that the bulk resistivity follows a simple activation law with no apparent carrier lifetime variation. The high field bulk de~Haas\,--\,van Alphen oscillations\index{SmB$_6$!quantum oscillations} also present mysteries, both as to their occurrence and their similarity to the signals seen in the trivalent, metallic hexaborides. The large ac conductivity is equally puzzling. There are clearly effects extrinsic to the fundamental physics of SmB$_6$ seen, for example, in the sample dependence of the low temperature specific heat and magnetic susceptibility. The source of the bulk low-temperature tail in the magnetic susceptibility needs to be identified as well as the bump seen in the specific heat at low temperature. One route to sorting out these puzzles will be a more detailed sample characterization of the crystals actually used in each experiment. Fully understanding the physics of defects appears to be a requisite for understanding this unusual compound. Surely there are other materials to be discovered carrying the same physics, and identifying the common physics can provide another path forward. YbB$_{12}$ seems a possible but not certain candidate, but other examples are highly desired and await discovery.

\vspace{-2pt}\section*{Acknowledgments}
\addcontentsline{toc}{section}{Acknowledgments}

The authors of this chapter acknowledge stimulating discussions with Sean M. Thomas, Joe D. Thompson, Steffen Wirth, Pascoal Pagliuso, Yun Suk Eo, Wesley Fuhrman, Cagliyan Kurdak, James Allen, Liang Fu, Piers Coleman, and Peter Riseborough. Work at Los Alamos was performed under the auspices of the U.\,S. Department of Energy, Office of Basic Energy Sciences, Division of Materials Science and Engineering.



\end{bibunit} 




\end{document}